\documentclass[preprint,12pt]{elsarticle}

\newcounter{bla}

\usepackage{lipsum}
\makeatletter
\def\ps@pprintTitle{%
 \let\@oddhead\@empty
 \let\@evenhead\@empty
 \def\@oddfoot{}%
 \let\@evenfoot\@oddfoot}
\makeatother

\usepackage[utf8]{inputenc}
\usepackage{textcomp}
\usepackage{bm}
\usepackage{natbib}
\usepackage{exscale}
\usepackage{siunitx}
\usepackage{graphicx}
\usepackage{amssymb,amsthm,amsmath}
\usepackage[version=4]{mhchem}
\usepackage{siunitx}
\usepackage{longtable,tabularx}
\usepackage{float}
\usepackage{booktabs}
\usepackage{listings} \usepackage{tikz} \usepackage{subfig}

\usepackage{soul}
\usepackage[ruled,vlined]{algorithm2e}

\usepackage{hyperref}

\makeatletter
\renewcommand*\env@matrix[1][\arraystretch]{%
  \edef\arraystretch{#1}%
  \hskip -\arraycolsep
  \let\@ifnextchar\new@ifnextchar
  \array{*\c@MaxMatrixCols c}}
\makeatother

\begin{document}

\begin{frontmatter}

\title{An efficient reconstruction algorithm for diffusion on
  triangular grids using the nodal discontinuous Galerkin method}

\author[mymainaddress]{Yang Song}

\author[mymainaddress]{Bhuvana
  Srinivasan\corref{mycorrespondingauthor}}
\cortext[mycorrespondingauthor]{Corresponding author}
\ead[url]{srinbhu@vt.edu}

\address[mymainaddress]{Kevin T. Crofton Department of Aerospace and
  Ocean Engineering, Virginia Tech, Blacksburg, VA 24060}

\begin{abstract}
High-energy-density (HED) hydrodynamics studies such as those relevant
to inertial confinement fusion and astrophysics require highly
disparate densities, temperatures, viscosities, and other diffusion
parameters over relatively short spatial scales.  This presents a
challenge for high-order accurate methods to effectively resolve the
hydrodynamics at these scales, particularly in the presence of highly
disparate diffusion.  A significant volume of engineering and physics
applications use an unstructured discontinuous Galerkin (DG) method
developed based on the finite element mesh generation and algorithmic
framework.  This work discusses the application of an affine
reconstructed nodal DG method for unstructured grids of triangles.
Solving the diffusion terms in the DG method is non-trivial due to the
solution representations being piecewise continuous.  Hence, the
diffusive flux is not defined on the interface of elements.  The
proposed numerical approach reconstructs a smooth solution in a
parallelogram that is enclosed by the quadrilateral formed by two
adjacent triangle elements. The interface between these two triangles
is the diagonal of the enclosed parallelogram. Similar to triangles,
the mapping of parallelograms from a physical domain to a reference
domain is an affine mapping, which is necessary for an accurate and
efficient implementation of the numerical algorithm.  Thus, all
computations can still be performed on the reference domain, which
promotes efficiency in computation and storage.  This reconstruction
does not make assumptions on choice of polynomial basis. Reconstructed
DG algorithms have previously been developed for modal implementations
of the convection-diffusion equations.  However, to the best of the
authors' knowledge, this is the first practical guideline that has
been proposed for applying the reconstructed algorithm on a nodal
discontinuous Galerkin method with a focus on accuracy and efficiency.
The algorithm is demonstrated on a number of benchmark cases as well
as a challenging substantive problem in HED hydrodynamics with highly
disparate diffusion parameters.
\end{abstract}

\begin{keyword}
nodal discontinuous Galerkin method; reconstruction; convection
diffusion equation; computational efficiency; unstructured; triangle
elements; high-energy-density hydrodynamics
\end{keyword}

\end{frontmatter}


\newcommand{\Transpose}{^{\mathsf{T}}}

\section{Introduction}
A number of problems of interest in physics and engineering, such as
those in fluid dynamics including high-energy-density hydrodynamics,
rely on geometric flexibility and randomized grid errors so the choice
of mesh does not impact the physics.  Hence, an unstructured nodal
discontinuous Galerkin (DG) scheme is utilized in this work ensuring
geometric flexibility along with high-order accuracy
\citep{NDGbook_2007}.  High-energy-density hydrodynamics studies such
as those relevant to inertial confinement fusion and astrophysics
require highly disparate densities, temperatures, viscosities, and
other diffusion parameters over relatively short spatial scales
\cite{clark2016three, srinivasan2014mitigating}.  This presents a
challenge for high-order accurate methods to effectively resolve the
hydrodynamics at these scales, particularly in the presence of highly
disparate diffusion.  This work provides the first practical guideline
on an accurate and efficient reconstructed algorithm for diffusion
using the nodal DG method on triangular elements with potential broad
impact on the large community of nodal DG applications using the
finite element mesh generation and algorithmic framework.

In recent years, the DG method has been successfully applied to
hyperbolic conservation laws \citep{bassi1997high, cockburn1998runge,
  srinivasan2010numerical, srinivasan2011numerical,
  cockburn2012discontinuous}. Due to its compactness, high order
accuracy, and versatility, the DG algorithm is favorable for
applications to convection-diffusion problems,
\begin{equation}
  \frac{\partial \bm{u}}{\partial t} + \nabla \cdot (\vec{v} \bm{u}) -
  \nabla \cdot (D \nabla \bm{u}) = \bm{s}
  \label{equ:convection_diffusion}
\end{equation}
where $\bm{u}$ represents conservative variables, $\vec{v}$ is the
velocity field, $D$ is the diffusion coefficient and $\bm{s}$
represents source terms. A significant amount of literature exists on
accurate and efficient DG implementations for the convection terms.

However, solving the diffusion term in DG is non-trivial. The
diffusive flux is not defined on the interface of elements as DG
solution representations are only piecewise continuous. Approximating
the diffusive flux as a simple arithmetic mean from both sides of the
interface is not appropriate as it ignores the possible jump of the
solutions.  A number of numerical algorithms have been proposed in the
DG community to approximate the diffusion operator with high order
accuracy, for example, Douglas and Dupont \citep{douglas1976interior},
Arnold \citep{arnold1982interior}, Cockburn and Shu
\citep{cockburn1998local}, Peraire and Persson
\citep{peraire2008compact}, Liu and Yan \citep{liu2009direct}, and
others.  However, all the above methods require large computational
effort relative to the algorithm presented here.

In 2005, Van Leer proposed a recovery-based DG algorithm to solve the
diffusion operator, where a new polynomial that is smoothly defined
across two adjacent elements is recovered from the two original
polynomials with order of $P$ \citep{vanLeer_2005}. The new polynomial
is of order $2P+1$ and is indistinguishable from the original
solutions defined across two cells in a weak sense. This
recovery-based method is a more natural and accurate way of
calculating the diffusive flux. This algorithm is further developed
and applied on a two dimensional structured mesh
\citep{nourgaliev2010recovery}.  However, the accuracy of the scheme
is affected not only by the diffusive part but also the hyperbolic
parts in the system. In fact, the order of accuracy is determined by
the least accurate component in the system. Hence, a highly accurate
diffusion solver does not increase the overall accuracy of the scheme
in solving convection-diffusion problems. Also, constructing an
appropriate basis function defined on the combination of two elements
is an involved process. More recently, a reconstruction-based DG
algorithm using Taylor basis functions is proposed in
\citep{Luo_2010}.  In this algorithm, similar to the recovery DG
algorithm, a smooth solution is reconstructed across two adjacent
elements. Unlike the recovery DG algorithm, the reconstructed solution
has the same polynomial order as the original solutions and is not
indistinguishable from the original solutions in a weak sense. The
reconstruction-based DG algorithm can solve the diffusion term with
the same order of accuracy as the hyperbolic solver, making the scheme
computationally efficient. Also, since the reconstructed polynomial
has the same order as the underlying DG solution, it is not necessary
to carefully construct a basis function that is well conditioned
across two elements. The choice of Taylor basis simplifies the
reconstruction process significantly although it suffers from
ill-conditioning.

Storage management and computational efficiency are playing
increasingly significant roles in modern computational software
especially for large-scale high fidelity simulations. Conventional DG
algorithms solve hyperbolic terms on a reference element, then
transform the solution to physical elements.  There are advantages
with respect to computational efficiency and memory management if the
reconstructed DG algorithm could be solved on a reference domain.
Depending on the shape of the elements (triangle, quadrilateral,
etc.), different memory requirements are dictated by the need to store
the transformation Jacobians.  Without careful treatment, this could
result in higher cost of either memory or computation for recovery or
reconstruction methods.  Thus, solving the diffusion operator using DG
in a stable, efficient, and accurate manner is still an open
question. It is worth mentioning that recent developments have been
made in the reconstructed DG algorithm to couple the direct DG method
\citep{yang2018reconstructed} with a first-order hyperbolic system
(FOHS) \citep{lou2018reconstructed}.  However, the primary focus of
this paper is on memory and computational efficiency while solving the
diffusion term. What is more, there is no guideline currently
available on how to apply the reconstruction technique directly on a
nodal DG method.  This work proposes a new reconstructed DG method
that is both storage- and computationally-efficient, and couples
naturally with the widely-used nodal DG algorithm described by
Hesthaven and Warburton\citep{NDGbook_2007}.  This algorithm ensures
that the reconstruction is performed on affine elements, where the
transformation Jacobian is constant inside an element. This
significantly reduces the storage (or computation) required for the
transformation Jacobians compared to non-affine elements.  This
algorithm is designed for unstructured meshes. Unstructured mesh is
known for producing random grid errors as opposed to the preferential
errors of a Cartesian mesh. This can be very important for certain
applications where complex or general geometries are involved.  A
challenging problem from high-energy-density hydrodynamics, with
highly disparate diffusion parameters over relatively short spatial
scales, is demonstrated in Section \ref{sec:ICF} using this novel
reconstruction nodal DG algorithm with unstructured meshes.

\section{Governing equation and discretization}

\subsection{Governing equation}
This work focuses on solving the diffusion operator
using a reconstructed DG method. The governing equation is the
diffusion equation,
\begin{equation}
  \frac{\partial u}{\partial t} = \nabla \cdot (D\nabla  u)
  \label{equ:diffusion}
\end{equation}
where D is the diffusion coefficient. Without losing generality, 
$D$ is assumed to be a positive constant in space and time.

\subsection{Discretization}
In DG, the numerical solution can be expressed as a direct sum of
local piecewise polynomials as
\begin{equation}
  u(\bm{x},t) \simeq u_h(\bm{x},t) = \bigoplus_{k=1}^K u_h^k(\bm{x},t) .
  \label{equ:DG_discretization}
\end{equation}

Replacing $u$ in equation \ref{equ:diffusion} with $u_h$ and
multiplying a test function $\phi_i$ and integrating over
non-overlapping cells $\Omega_k$, where $k = 1,...,K$, will give a
typical DG treatment,
\begin{equation}
  \int_{\Omega_k}\left(\frac{\partial u_h^k}{\partial t}\phi_i^k -
  D(\nabla^2  u_h^k)\phi_i^k \right) d{\Omega} =0 .
  \label{equ:DG_diffusion}
\end{equation}

A DG scheme can be obtained by integrating the second term in equation
\ref{equ:DG_diffusion} by parts,

\begin{equation}
  \int_{\Omega_k}\left(\frac{\partial u_h^k}{\partial t}\phi_i^k + D
  \nabla u_h^k \cdot \nabla \phi_i^k \right)d\Omega -
  D\int_{\partial\Omega_k}\left(\phi_i^k \bm{\hat{n}} \cdot {\nabla
    \tilde{u}^k} \right)d{\partial\Omega} =0 .
  \label{equ:DG_diffusion_once}
\end{equation}

Since $u^k_h$ is discontinuous at the cell interface, the diffusive
flux $\nabla u_h^k$ in the surface integration is not directly
available on the boundary of $\Omega_k$ and cannot be treated as an
advective flux, thus it cannot be simply approximated by a Riemann
flux solver \citep{vanLeer_2007, Luo_2010}. Hence, $\nabla u_h^k$ is
replaced by a reconstructed solution $\nabla \tilde{u}^k$ that is
smoothly defined at the interface. The details of this reconstruction
algorithm will be discussed in section \ref{sec:affinereconstruction}.

\section{Nodal discontinuous Galerkin method} \label{sec:NDG method}
Following the nodal DG algorithm from \cite{NDGbook_2007}, the test
function and basis function are chosen to be Lagrange polynomials,
$\ell_i$. For the sake of simplicity, the subscript $h$ is dropped from
now on. Then equation \ref{equ:DG_diffusion_once} can be rewritten
as
\begin{equation}
  \int_{\Omega_k}\left(\frac{\partial u^k}{\partial t}\ell_i^k + D
  \nabla u^k \cdot \nabla \ell_i^k \right)d\Omega -
  D\int_{\partial\Omega_k}\left(\ell_i^k \bm{\hat{n}} \cdot {\nabla
    \tilde{u}^k} \right)d{\partial\Omega} =0 .
  \label{equ:NDG_diffusion_once}
\end{equation}
Solutions on Legendre-Gauss-Lobatto (LGL) nodes
\citep{abramowitz1972handbook} are chosen to be the expansion
coefficients. Assume the polynomial order is $P$ and $\bm{x}_j^k$ 
are the LGL nodes defined on $\Omega_k$, then the solution in
$\Omega_k$ can be represented as the nodal expansion
\begin{equation}
  u^k(\bm{x},t) = \sum^{N_p}_{j=1} u^k(\bm{x}_j^k,t) \ell_j^k(\bm{x}) ,
\end{equation}
where $N_p = (P+1)(P+2)/2$ is the total number of nodes or unknowns in
$\Omega_k$ and $\bm{u}^k = [u^k(\bm{x}^k_1,t) ,\dots,
  u^k(\bm{x}^k_{N_p},t)]\Transpose$. The modal
expansion of the solution is introduced,
\begin{equation}
  u^k(\bm{x},t) = \sum^{N_p}_{j=1} \hat{u}_j^k(t) \psi_j^k(\bm{x}) ,
\end{equation}
where $\hat{\bm{u}}^k = [\hat{u}^k_1(t)
  ,\dots,\hat{u}^k_{N_p}(t)]\Transpose$ are the modal expansion
coefficients and $\psi_j^k(\bm{x})$ are the orthonormal modal
polynomial basis in $\Omega_k$. For more details of how to construct
$\psi_j$ in triangular element, please refer to \cite{NDGbook_2007}. 
The Vandermonde matrix $\mathcal{V}^k$ is defined as
\begin{equation}
  \mathcal{V}_{ij}^k = \psi_j^k(\bm{x_i}) ,
\end{equation}
such that
\begin{equation}
  \bm{u}^k = \mathcal{V}^k \hat{\bm{u}}^k .
\label{equ:Vandermonde}
\end{equation}

\begin{figure}[!htb]
 \centering
 \includegraphics[width=0.6\linewidth]{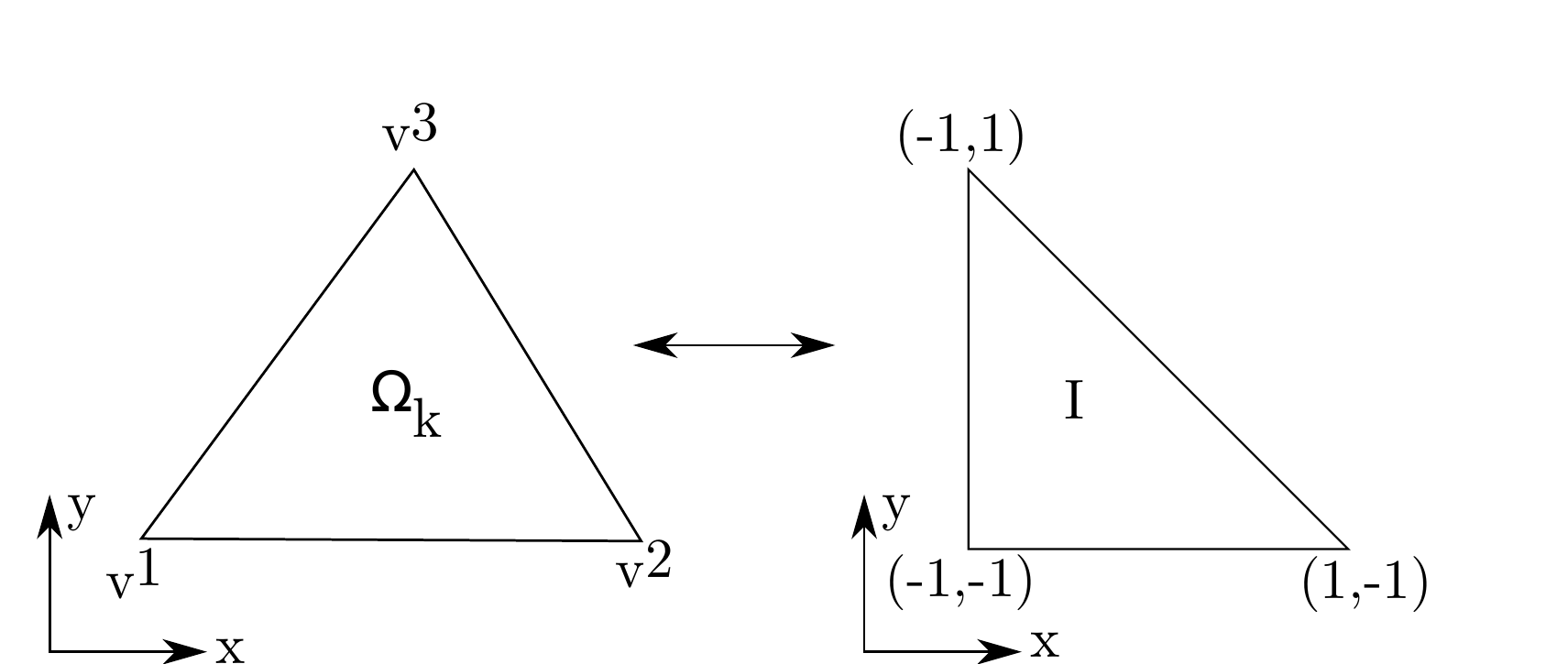}
 \caption{Affine transformation between physical element $\Omega_k$
   and reference element $I$}
  \label{fig:triangle_map}
\end{figure}

In the nodal DG method \cite{NDGbook_2007}, all
computations can be performed on the reference triangle $I =
\{\bm{r}=(r,s) | (r,s) \geq -1; r+s \leq 0 \}$.  Since the mapping for
triangular elements is an affine transformation \citep{veblen1918projective, berger1987geometry}, the Jacobians of this
mapping are constant in a triangle. This mapping is shown in Figure
\ref{fig:triangle_map} and described in equations \ref{equ:mapping_tri}
and \ref{equ:mapping_tri_diff},

\begin{equation}
  \bm{x} = -\frac{r+s}{2}\bm{v}^1 + \frac{r+1}{2}\bm{v}^2+
  \frac{s+1}{2}\bm{v}^3 ,
  \label{equ:mapping_tri}
\end{equation}

\begin{equation}
  \begin{aligned} 
    & (x_r,y_r) = \frac{\bm{v}^2 - \bm{v}^1}{2} & \text{, } (x_s,y_s)
    = \frac{\bm{v}^3 - \bm{v}^1}{2} .\\ 
  \end{aligned} 
  \label{equ:mapping_tri_diff}
\end{equation}

The Jacobians of this mapping are
described in equations \ref{equ:triangle_jacobian_a} and
\ref{equ:triangle_jacobian_b},
\begin{equation}
  r_x = \frac{y_s}{J} \text{, } r_y = -\frac{x_s}{J} \text{, } s_x =
  -\frac{y_r}{J} \text{, } s_y = \frac{x_r}{J} ,
  \label{equ:triangle_jacobian_a}
\end{equation}
\begin{equation}
  J = x_r y_s -x_s y_r .
  \label{equ:triangle_jacobian_b}
\end{equation}

For the remainder of this paper, any variable or matrix without the
element index superscript $k$ is defined on $I$. Now, equation
\ref{equ:NDG_diffusion_once} can be written as
\begin{equation}
  \frac{\partial \bm{u}^k}{\partial t} + D \left(
       {M^k}^{-1}{\bm{S}^k}\Transpose \cdot \nabla \bm{u}^k \right) -
       D \sum_{f=1}^3 \textrm{LIFT}_f^k \left( \hat{\bm{n}}^k_f\cdot
       \nabla\tilde{\bm{u}}^k_f\right) =0 ,
  \label{equ:NDG_diffusion_once_matrixForm}
\end{equation}
where the mass matrix and stiffness matrix are defined as

\begin{equation}
M_{ij}^k = \int_{\Omega_k}\ell_i^k \ell_j^k d\Omega = 
J^k\int_{I}\ell_i \ell_j d I = J^k M ,
\end{equation} 
\begin{equation}
\begin{aligned}
\bm{S}_{ij}^k &=\int_{\Omega_k}\ell_i^k \nabla \ell_j^k d\Omega  \\
&=J^k \int_{I}\ell_i
\begin{bmatrix}
r_x & s_x\\
r_y & s_y\\
\end{bmatrix}^k 
\begin{bmatrix}[1.5]
\frac{\partial \ell_j}{\partial r} \\
\frac{\partial \ell_j}{\partial s}
\end{bmatrix} d I \\
&= J^k\bm{r}_{\bm{x}}^k \int_{I}\ell_i \nabla \ell_j d I \\
&= J^k \bm{r}_{\bm{x}}^k \bm{S},
\end{aligned} 
\end{equation}
respectively. Only reference mass, stiffness matrices, and geometric
factors need to be stored. The lift operator is defined as
\begin{equation}
\textrm{LIFT}_f^k \left( \hat{\bm{n}}^k_f\cdot \nabla\tilde{\bm{u}}^k_f\right) 
= {M^k}^{-1} \int_{\partial \Omega_k^f} \ell_i^k \hat{\bm{n}}^k_f
\cdot \nabla \tilde{u}^k_f d\partial\Omega. 
\label{lift_physicalDomain}
\end{equation}
Here, the surface integration cannot be easily transformed to the
reference domain, as the reconstructed element is not guaranteed to
share the same mapping transformation of triangular elements as
described in equation \ref{equ:mapping_tri} and
\ref{equ:mapping_tri_diff}. This means that this surface integration
needs to be pre-calculated and stored on all elements, which is
computationally inefficient. This will be discussed in the following
section.

 Now, equation
\ref{equ:NDG_diffusion_once_matrixForm} can be written as
\begin{equation}
  \frac{\partial \bm{u}^k}{\partial t} + D \left(
       {M}^{-1}{\bm{S}}\Transpose \cdot \nabla \bm{u}^k \right) - D
       \sum_{f=1}^3 \textrm{LIFT}_f^k \left( \hat{\bm{n}}^k_f\cdot \nabla\tilde{\bm{u}}^k_f\right) =0 .
  \label{equ:NDG_diffusion_once_matrixForm_reference}
\end{equation}
If $D$ is not a constant, but a function of space and time, and also
not isotropic (i.e. $\bm{D} = (D_x, D_y)$) then equation
\ref{equ:NDG_diffusion_once_matrixForm_reference} can be rewritten as
\begin{equation}
  \frac{\partial \bm{u}^k}{\partial t} +  \left(
       {M}^{-1}{\bm{S}}\Transpose \cdot \bm{D} \nabla \bm{u}^k \right) - 
       \sum_{f=1}^3 \textrm{LIFT}_f^k \left(\hat{\bm{n}}^k_f\cdot
       \widetilde{\bm{D} \nabla\tilde{\bm{u}}_f}\right) =0 .
  \label{equ:NDG_diffusion_once_matrixForm_reference_general}
\end{equation} 
This is summarized using the pseudo-code in Algorithm
\ref{algo:nonconstDiff}, where the algorithmic details for the novel
reconstruction method (RDG) are described in section
\ref{sec:reconstruction} with the corresponding pseudocode in
Algorithm \ref{algo:rdg}.

\begin{algorithm}[H]
  \SetAlgoLined
  \tcp{$\bm{u}^1$ and $\bm{u}^2$ are nodal DG solutions on two adjacent elements $\Omega_1$ and $\Omega_2$}
  \tcp{Perform the reconstruction of $u$ and get the reconstructed solution $\tilde{u}$ on the reconstructed element which consists of two triangles}
  $\tilde{\bm{u}}$  = RDG($\bm{u}^1$, $\bm{u}^2$);

  \tcp{Project $\tilde{\bm{u}}$ to the two triangles that forms the reconstructed element}
  [$\tilde{\bm{u}}^1$,  $\tilde{\bm{u}}^2$] = Separation($\tilde{\bm{u}}$);

  \tcp{Project $\bm{D}$ to the two triangles that forms the reconstructed element}
  $\bm{D}_p^1$ = $\mathcal{V}_{p1} \bm{D}^1$;

  $\bm{D}_p^2$ = $\mathcal{V}_{p2} \bm{D}^2$;
  
  \tcp{Calculate the gradients of $\tilde{u}$}
  $\nabla \tilde{\bm{u}}^1$ = Grad($\tilde{\bm{u}}^1$);
  
  $\nabla \tilde{\bm{u}}^2$ = Grad($\tilde{\bm{u}}^2$);
  
  \tcp{Perform the reconstruction}
  $\widetilde{\bm{D} \nabla\tilde{\bm{u}}}$ = RDG($\bm{D}_p^1 \nabla\tilde{\bm{u}}^1$, $\bm{D}_p^2 \nabla\tilde{\bm{u}}^2$)
  \caption{Reconstruction algorithm for non-constant diffusion coefficients using equation \ref{equ:NDG_diffusion_once_matrixForm_reference_general}} \label{algo:nonconstDiff}
\end{algorithm}

An alternative way of calculating the reconstructed solution for the
surface term is,
\begin{equation}
  \frac{\partial \bm{u}^k}{\partial t} +  \left(
       {M}^{-1}{\bm{S}}\Transpose \cdot \bm{D} \nabla \bm{u}^k \right) - 
       \sum_{f=1}^3 \textrm{LIFT}_f^k \left(\hat{\bm{n}}^k_f\cdot
       \tilde{\bm{D}} \nabla\tilde{\bm{u}}_f\right) =0 .
  \label{equ:NDG_diffusion_once_matrixForm_reference_general_alternative}
\end{equation}
This is summarized in the pseudo-code in Algorithm
\ref{algo:nonconstDiff2}.

\begin{algorithm}[H]
  \SetAlgoLined
  \tcp{Perform the reconstruction of $u$ and get the reconstructed solution $\tilde{u}$ on the reconstructed element which consists of two triangles}
  $\tilde{\bm{u}}$  = RDG($\bm{u}^1$, $\bm{u}^2$);

  \tcp{Calculate the gradients of $\tilde{u}$}
  $\nabla \tilde{\bm{u}}$ = Grad($\tilde{\bm{u}}$);
  
  \tcp{Perform the reconstruction of $\bm{D}$ and get the reconstructed solution $\tilde{\bm{D}}$ on the reconstructed element which consists of two triangles}
  $\tilde{\bm{D}}$  = RDG($\bm{D}^1$, $\bm{D}^2$);

  \tcp{Calculate the product}
  $\tilde{\bm{D}} \nabla\tilde{\bm{u}}$  = Product($\tilde{\bm{D}}$, $\nabla\tilde{\bm{u}}$)
  \caption{Reconstruction algorithm for non-constant diffusion coefficients using equation \ref{equ:NDG_diffusion_once_matrixForm_reference_general_alternative}} \label{algo:nonconstDiff2}
\end{algorithm}

Test results indicate minimal differences between the two
reconstructed formulations described in equations
\ref{equ:NDG_diffusion_once_matrixForm_reference_general} and
\ref{equ:NDG_diffusion_once_matrixForm_reference_general_alternative}.

\section{Affine reconstructed algorithm} \label{sec:affinereconstruction}

The use of affine elements in the reconstruction makes the memory
storage and computation more efficient as it avoids the higher order
transformation function in the reconstructed element. The proposed
method is designed for arbitrary mesh type including unstructured
meshes which are known to have randomized grid errors.  This section
describes the motivation and details for an affine reconstructed DG
(aRDG) algorithm.

\subsection{Non-affine mapping in quadrilaterals}

\begin{figure}[!htb]
  \centering 
  \subfloat[LGL nodes on $I^q$]{\includegraphics[width=0.33\linewidth]{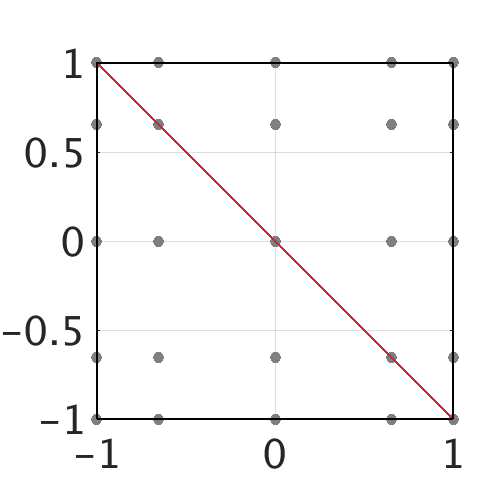}}
  \subfloat[$\Omega^q_{1}$]{\includegraphics[width=0.33\linewidth]{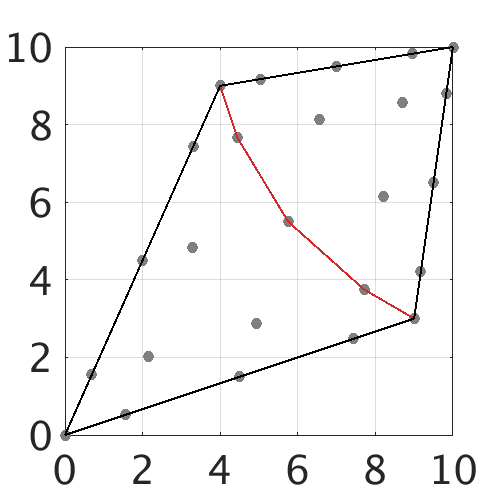}}
  \subfloat[$\Omega^q_{2}$]{\includegraphics[width=0.33\linewidth]{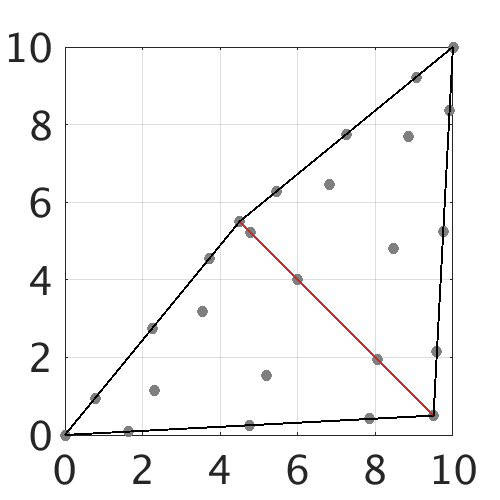}}
  \caption{Mapping transformation in quadrilaterals. (a) tensor
    product of LGL nodes on $I^q$; (b) transformation from $I^q$ to
    $\Omega^q_{1}$ that has a curved diagonal; (c) transformation from
    $I^q$ to $\Omega^q_{2}$ that has a straight diagonal but with
    asymmetric nodes along the diagonal.}
  \label{quad_mapping_distort}
\end{figure}

To obtain a reconstructed solution that is smoothly defined at the
interface, the reconstruction needs to be performed on the combination
of two triangles, which is a quadrilateral.  Hence, it is important to
consider the mapping transformation between a quadrilateral element
$\Omega^q$ and a reference square element $I^q = \{\bm{R}=(R,S) | -1
\leq (R,S) \leq 1 \}$. Here superscript $q$ refers to quadrilateral.
This mapping is described in equation \ref{equ:mapping_quad},
\begin{equation}
  \bm{X} = \frac{1}{4}(1-R)(1-S)\bm{v}^1
  +\frac{1}{4}(1+R)(1-S)\bm{v}^2 +\frac{1}{4}(1+R)(1+S)\bm{v}^3
  +\frac{1}{4}(1-R)(1+S)\bm{v}^4
  \label{equ:mapping_quad},
\end{equation}
which is not always an affine mapping. Thus, assuming $I(\bm{r})$ for
the reference triangle and $I^q(\bm{R})$ for the reference square
element share the same coordinate system, then $\Omega(\bm{x})$ and
$\Omega^q(\bm{X})$ are not in the same physical coordinate system. To
demonstrate this, $P4$ ($Pn$ denotes polynomial order $n$) tensor
product nodal points in $I^q$, as shown in Figure
\ref{quad_mapping_distort}-a, are mapped to an arbitrary quadrilateral
element $\Omega^q_{1}$ through equation \ref{equ:mapping_quad}, as
shown in Figure \ref{quad_mapping_distort}-b.  Note that the nodes on
the diagonal of $\Omega^q_{1}$ are curved and do not represent the
straight interface between the two triangles. Figure
\ref{quad_mapping_distort}-c provides another example where the
diagonal of the quadrilateral in $\Omega^q_{2}$ is not curved but the
nodes on diagonal are not symmetric. This shows that the
diagonal of $\Omega^q$ does not represent the interface between two
triangular elements. This makes the reconstruction unfavorable as the
surface integration described in equation \ref{lift_physicalDomain}
can then only be evaluated on the physical domain, which is
inefficient for both computation and storage management.

\subsection{Enclosed parallelogram}

\begin{figure}[!htb]
  \centering
  \includegraphics[width=0.33\linewidth]{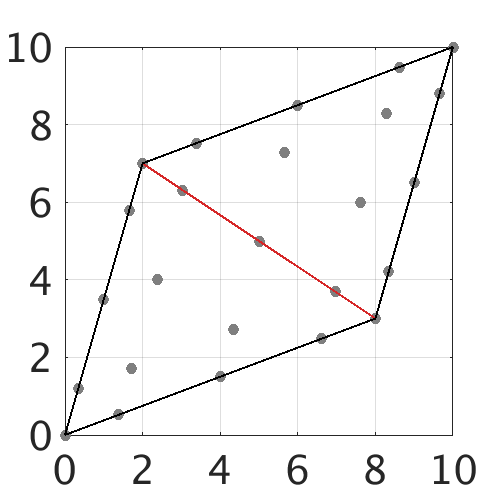}
  \caption{Tensor product of LGL nodes on a parallelogram formed by
    two adjacent triangles.}
  \label{parallelogram}
\end{figure}
\begin{figure}[!htb]
  \centering \subfloat[ two
    triangles]{\includegraphics[width=0.33\linewidth]{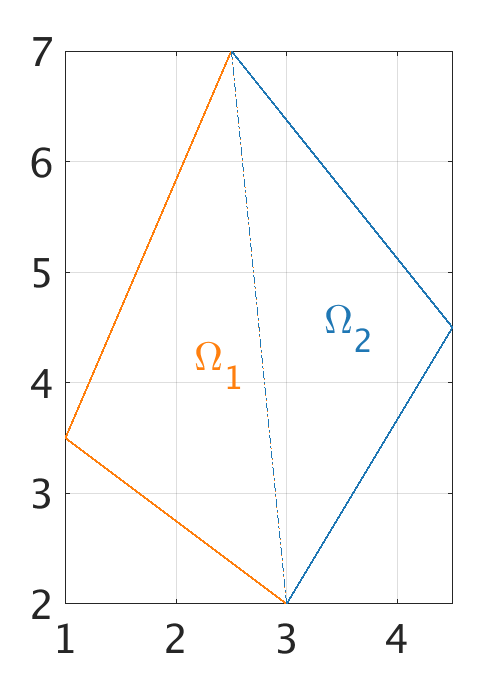}}
  \subfloat[enclosed
    parallelogram]{\includegraphics[width=0.33\linewidth]{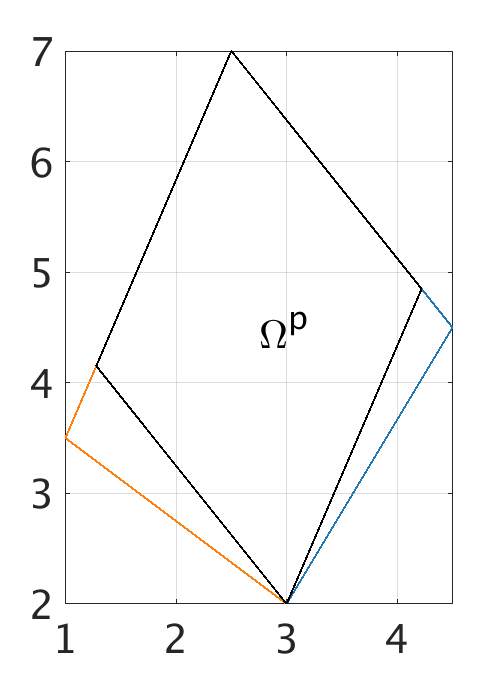}}
  \subfloat[two triangles from the enclosed parallelogram
  ]{\includegraphics[width=0.33\linewidth]{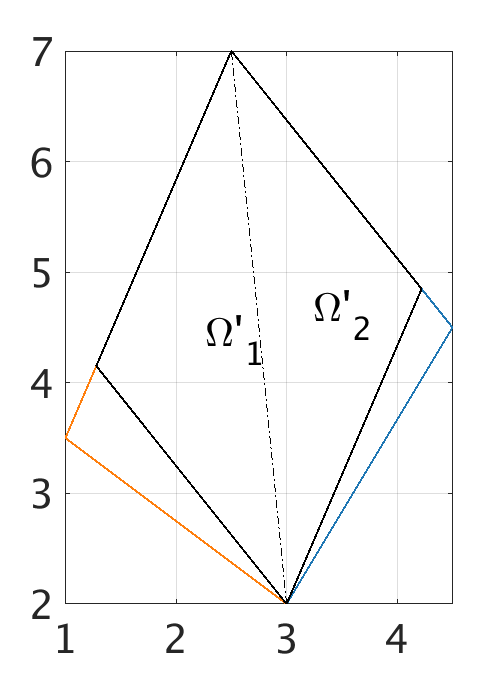}}
  \caption{Illustrations of an enclosed parallelogram found in two
    adjacent triangles.}
  \label{enclosedParallelogram_reconstruction}
\end{figure}

The mapping from equation \ref{equ:mapping_quad} can be reduced to
affine mapping when the physical quadrilateral $\Omega^q$ is a
parallelogram, which is shown in Figure \ref{parallelogram}.  For any
quadrilateral $\Omega^q$ formed by two adjacent triangles $\Omega_1$
and $\Omega_2$, one can always find an enclosed parallelogram
$\Omega^p$ that shares the same diagonal with $\Omega^q$, which is
also the interface between two triangles. This is demonstrated in
Figure \ref{enclosedParallelogram_reconstruction}. Once $\Omega^p$ is
found, the solution from $\Omega_1$ and $\Omega_2$ is projected onto
the two smaller triangles $\Omega'_1$ and $\Omega'_2$ that form the
parallelogram. Then the solution from these two triangles can be used
to reconstruct a polynomial $\tilde{u}$ that is continuously defined
in the parallelogram.  This reconstruction can be done in the logical
element $I^q = I + I^{-1}$, where $I^{-1} = \{\bm{r}=(r,s) | (r,s)
\leq 1; r+s \geq 0 \}$, with solution of $\Omega'_1$ projected on $I$
and solution $\Omega'_2$ projected on $I^{-1}$, when the shared
interface in $\Omega'_1$ and $\Omega'_2$ is the hypotenuse in $I$ and
$I_q$. This is because the nodes on the diagonal of $\Omega^p$ are
located exactly at the nodes on the interface of $\Omega'_1$ and
$\Omega'_2$. In other words, the mapping transformation between
$\Omega^p$ and $I^q$ is identical to the mapping transformation
between $\Omega'$ and $I$. The formula for the projection is provided
here but the reconstruction procedure will be discussed in detail in
section \ref{sec:reconstruction}. Once the new vertices are found for
$\Omega'_1$ and $\Omega'_2$, one can easily construct a projection
Vandermonde matrix $\mathcal{V}_p$ that projects the modal expansion
coefficients $\bm{\hat{u}}$ on $\Omega$ to the nodal solution
$\bm{u}'$ on $\Omega'$, as described in equation
\ref{equ:NodalToModalProjection},

\begin{equation}
  \bm{u}' = \mathcal{V}_p \bm{\hat{u}} .
  \label{equ:NodalToModalProjection}
\end{equation}
Now equation \ref{lift_physicalDomain} can be rewritten as
\begin{equation}
\begin{aligned}
\textrm{LIFT}_f^k \left( \hat{\bm{n}}^k_f\cdot \nabla\tilde{\bm{u}}^k_f\right) 
&= {M^k}^{-1} \int_{\partial \Omega_k^f} \ell_i^k \hat{\bm{n}}^k_f
\cdot \nabla \tilde{u}^k_f d\partial\Omega \\ 
&={J^k  M}^{-1} \left(
\int_{\partial \Omega_k^f} \ell_i^k \tilde{\ell}_r^{k,f} d\partial\Omega \right)
\hat{\bm{n}}^k_f \cdot \nabla \tilde{\bm{u}}^k_f \\
 &= \frac{J^k_f}{J^k}{M}^{-1}  \left( \int_{\partial I^f}
\ell_i \ell_r^f d\partial I \right) \hat{\bm{n}}^k_f \cdot  
\nabla \tilde{\bm{u}}^k_f\\
&= \frac{J^k_f}{J^k}\textrm{LIFT}_f \left(\hat{\bm{n}}^k_f\cdot \nabla \tilde{\bm{u}}^k_f\right),
\end{aligned} 
\label{lift_physicalDomain_ardg}
\end{equation}
where $\tilde{\ell}_r^{k,f}$ is the basis function defined on the
diagonal of the reconstructed enclosed parallelogram element, which is
the same as the basis function defined on the edge of the
triangle. $\ell_r^f$ is the basis function defined on edge $f$ in $I$.
$J^k_f$ is the transformation Jacobian along edge $f$ of
$\Omega_k$. $J^k_f$ can also be seen as the ratio between the length
of $\Omega_k^f$ and $I^f$. $\nabla \tilde{\bm{u}}^k_{f=1,2,3} $
are $N_{fp} \times 1$ arrays of the gradients of the reconstructed
nodal solutions on the three edges of element $\Omega_k$. $N_{fp} =
P+1$ is the total number of nodes on one edge. $\nabla
\tilde{u}^k_f(\bm{x})$ can be calculated as,
\begin{equation}
  \nabla \tilde{u}(\bm{x})^k_f = \begin{bmatrix}
r'_x & s'_x\\
r'_y & s'_y\\
\end{bmatrix}^{k,f}
\begin{bmatrix}[1.5]
\frac{\partial \tilde{u}^k_f}{\partial r} \\
\frac{\partial \tilde{u}^k_f}{\partial s}
\end{bmatrix},
\end{equation}
where the geometric factors are constant in a parallelogram, which
requires much less storage compared to quadrilateral elements.
Equation \ref{equ:NDG_diffusion_once_matrixForm_reference} now can be
written as,
\begin{equation}
  \frac{\partial \bm{u}^k}{\partial t} + D \left(
       {M}^{-1}{\bm{S}}\Transpose \cdot \nabla \bm{u}^k \right) - D
       \sum_{f=1}^3 \frac{J^k_f}{J^k}\textrm{LIFT}_f \left(\hat{\bm{n}}^k_f\cdot \nabla \tilde{\bm{u}}^k_f\right)=0,
  \label{equ:NDG_diffusion_once_matrixForm_Fullyreference}
\end{equation}
in which all matrices are defined in $I$.  This form has advantages
for numerical implementation as the matrices can be pre-calculated
while also using minimal storage.

\subsection{Storage and computational efficiency} \label{sec:storage}
The use of an affine transformation on the reconstructed element has
significant computational advantages. The geometric factors $J$ and
$\bm{R_X}$ are constant in an affine element, thus reducing the
storage requirement significantly for affine elements. The comparison
of the storage required for the geometric factors between
parallelogram elements and quadrilateral elements is presented in
Table \ref{table:ardg_storage}. This storage is required for each
interior edge of the mesh. The requirement for the mass matrices are
also tabulated in Table \ref{table:ardg_storage}. For parallelogram
elements, the computation can be performed on the reference domain
hence the mass matrix is only defined on the logical domain resulting
in significantly lower storage requirements.  For general
quadrilateral elements, however, the transformation is different from
that of triangular elements, hence the computation needs to be
performed on the physical domain requiring storage of the mass matrix
for each element.

  \begin{table}[!htbp]
    \centering
    \begin{tabular}{l|c|c}
      \toprule
      & Parallelogram & Quadrilateral \\
      \midrule
      \begin{tabular}{r}
        $P$   \\
        \midrule
        $N_{\mathrm{P}}$    \\
        {$J$} \\
        {$\bm{R_X}$} \\
        {$\mathcal{M}$}\\
      \end{tabular}  & 
      \begin{tabular}{l l l l l}
        1 & 2 & 3 &4 &5     \\
        \midrule
        4 & 9 & 16 &25 &36  \\
        1 & 1 & 1 & 1 & 1   \\
        4 & 4 & 4 & 4 & 4   \\
        16 & 81 &256  &625 &1296\\
      \end{tabular} &
      \begin{tabular}{l l l l l}
        1 & 2 & 3 &4 &5   \\
        \midrule
        4 & 9 & 16 &25 &36   \\
        4 & 9 & 16 &25 &36   \\
        16 & 36 & 64 & 100 & 144   \\
        16 & 81 &256  &625 &1296   \\
      \end{tabular} \\ 
      \bottomrule
    \end{tabular} 
    \caption{Comparison of storage requirements (values correspond to
      the number of values stored for the geometric factors, mass and
      stiffness matrices) between parallelogram and quadrilateral
      elements. The storage indicated for $J$ and $\bm{R_X}$ are
      required for each reconstructed element (each interior edge in
      the mesh). The storage indicated for $\mathcal{M}$ is required
      only for reference element if the computation can be performed
      on the reference element, and is required for each reconstructed
      element if the computation needs to be performed on the physical
      domain.}
\label{table:ardg_storage}
  \end{table}

\subsection{Reordering nodes in the reference domain $(r,s)$}
\begin{figure}[!htb]
  \centering \subfloat[LGL nodes in $I$ for $(\bm{v}^1,\bm{v}^2)$ as
    hypotenuse]{\includegraphics[width=0.33\linewidth]{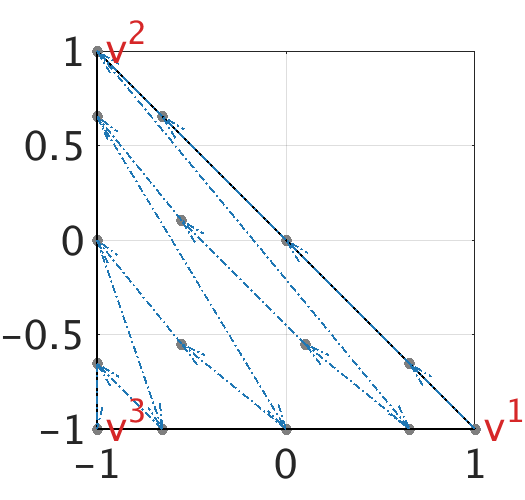}}
  \qquad \qquad \subfloat[LGL nodes in $I$ for $(\bm{v}^3,\bm{v}^1)$
    as
    hypotenuse]{\includegraphics[width=0.33\linewidth]{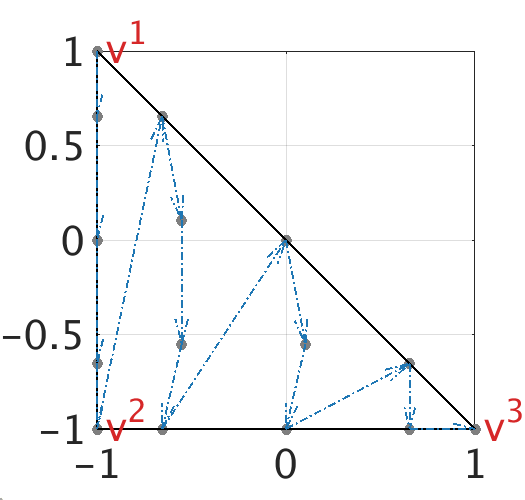}}
 \caption{Reordering of nodes in $I$ to map the hypotenuse
   $((1,-1),(-1,1))$ on (a) edge $(\bm{v}^1,\bm{v}^2)$ or (b) edge
   $(\bm{v}^3,\bm{v}^1)$ in $\Omega$. The ordering of nodes in
   $\Omega$ remains unchanged.}
  \label{fig:logical_ordering}
\end{figure}

Every edge of $\Omega_k$ that has a neighboring element will need to
be the hypotenuse in $I$ for the reconstruction. An immediate solution
to this would be changing the ordering of the vertices $[\bm{v}^1,
  \bm{v}^2, \bm{v}^3]$ in equation \ref{equ:mapping_tri} to change the
ordering of the nodes in $\Omega_k$, so that the target edge of
$\Omega_k$ can be remapped to the hypotenuse of $I$. However, this
needs to be done for two other edges of each element, and requires
either large computational effort if it is calculated during run-time
or duplicated large storage if it is pre-calculated. This breaks the
simplicity and efficiency of this scheme.  A more efficient way to
solve this is to change the ordering of nodes in $I$ to map its
hypotenuse to the target edge in $\Omega$, without changing the
ordering of nodes in $\Omega$.  As this is in the
reference domain $(r,s)$, only two additional orderings of $(r,s)$
need to be pre-calculated and stored. That is $[(1,-1), (-1,1),
  (-1,-1)]$ if $(\bm{v}^1, \bm{v}^2)$ needs to be the hypotenuse in
$I$, and $[(-1,1), (-1,-1), (1,-1)]$ if $(\bm{v}^3, \bm{v}^1)$ needs to
be the hypotenuse in $I$. This is demonstrated in Figure
\ref{fig:logical_ordering}.

There are three orderings of $(r,s)$ in $I$ that can be used for
performing the aRDG treatment on three edges of $\Omega$. Accordingly,
three Vandermonde matrices $[\mathcal{V}_{r1}, \mathcal{V}_{r2},
  \mathcal{V}_{r3}]$ can be generated to project the original nodal
solutions $\bm{u}$ on $\Omega$ to modal expansion coefficients
$\bm{\hat{u}_f}$ on $I$ so the desired edge $f$ matches the
hypotenuse.  This is described as

\begin{equation}
  \bm{\hat{u}_f} = \mathcal{V}_{rf}^{-1} \bm{u}.
  \label{equ:changeOfHypotenuseProjection}
\end{equation}

Combining equations \ref{equ:Vandermonde},
\ref{equ:NodalToModalProjection}, and
\ref{equ:changeOfHypotenuseProjection}, the modal expansion
coefficient $\bm{\hat{u}_f}'$ is calculated in $\Omega'$, where the
edge $f$ in $\Omega$ (or $\Omega'$) is the hypotenuse in $I$, from the
nodal solution $\bm{u}$ in $\Omega$, as
\begin{equation}
  \bm{\hat{u}_f}' = \mathcal{V}^{-1}\mathcal{V}_p\mathcal{V}_{rf}^{-1}
  \bm{u} .
  \label{equ:totalProjection}
\end{equation}
This expression can also be precomputed using any symbolic solver.

\subsection{Reconstruction}  \label{sec:reconstruction}
The components necessary for the reconstruction have been described to
this point.  The reconstruction process is performed using the modal
solution, which is computed from the Vandermonde matrix and the nodal
solution in the two smaller triangles that form the enclosed
parallelogram. Similar to the recovery \cite{vanLeer_2005} and the
reconstruction \cite{Luo_2010} methods, a new polynomial is
constructed that is smoothly defined across two adjacent cells,
\begin{equation}
  \begin{aligned} 
    &\int_{\Omega'_1}\sum_{r=1}^{M_p}\tilde{\hat{u}}_r\tilde{\psi}_r\psi_md\Omega=
    \int_{\Omega'_1}\sum_{r=1}^{N_p}{\hat{u'}}_r^1\psi_r\psi_md\Omega ,\\ &\int_{\Omega'_2}\sum_{r=1}^{M_p}\tilde{\hat{u}}_r\tilde{\psi}_r\psi_md\Omega=
    \int_{\Omega'_2}\sum_{r=1}^{N_p}{\hat{u'}}_r^2\psi_r\psi_md\Omega ,\\
  \end{aligned} 
\end{equation}  
where $N_p = (P+1)(P+2)/2$ is the number of modes in a triangle and
$M_p$ is the number of modes in the parallelogram,
respectively. $\hat{u'}^1_r$ and $\hat{u'}^2_r$ are the modal
solutions on the two smaller triangles $\Omega_1$ and
$\Omega_2$. $\tilde{\hat{u}}_r$ is the reconstructed modal solution on
the parallelogram. Using tensor product of Gauss-Legendre polynomial
basis for the parallelogram, $M_p =(P+1)(P+1)$. This system has $2N_p$
equations and $M_p$ unknowns. This affine reconstruction method solves
$(P+1)^2$ unknowns from $(P+1)(P+2)$ equations which differs from the
$\frac{(P+1)(P+2)}{2}$ unknowns (potentially with additional higher
order correction terms) in the work of \citep{Luo_2010}. This system
is solved using a least-squares method described in \citep{Luo_2010}.
The aRDG algorithm can be summarized in the pseudo-code in Algorithm
\ref{algo:rdg}.

\begin{algorithm}[H]
  \SetAlgoLined
  \tcp{$\bm{u}^1$ and $\bm{u}^2$ are nodal DG solutions on two adjacent elements $\Omega_1$ and $\Omega_2$. $f^1$ and $f^2$ are the local face indices of the interface in these two elements}

  \tcp{Calculate the Vandermonde Matrices that project $\bm{u}^1$ and $\bm{u}^2$ to ${\bm{u}'}^1$ and ${\bm{u}'}^2$ on the enclosed triangles}
  $\mathcal{V}_{p1}$ = getProjectVandermonde($\Omega_1$);
    
  $\mathcal{V}_{p2}$ = getProjectVandermonde($\Omega_2$);
    
  \tcp{Calculate the Vandermonde matrices that rotate ${\bm{u}'}^1$ and ${\bm{u}'}^2$, so that the interface is on the hypotenuse of $\Omega_1$ and $\Omega_2$ in the reference domain}  
  $\mathcal{V}_{r1}$ = getRotateVandermonde($f^1$);

  $\mathcal{V}_{r2}$ = getRotateVandermonde($f^2$);
  
  \tcp{Calculate the rotated modal solution on the two enclosed triangles}
  $\hat{\bm{u}}^1 = \mathcal{V}^{-1}\mathcal{V}_{p1}\mathcal{V}_{r1}^{-1} \bm{u}^1$;

  $\hat{\bm{u}}^2 = \mathcal{V}^{-1}\mathcal{V}_{p2}\mathcal{V}_{r2}^{-1} \bm{u}^2$;

  \tcp{Perform the reconstruction}
  $\tilde{\hat{u}} = modalReconstruction(\hat{\bm{u}}^1, \hat{\bm{u}}^2)$;
  \caption{aRDG algorithm} \label{algo:rdg}
\end{algorithm}

\section{Results} \label{sec:results}

\begin{figure}[!htb]
  \subfloat[]{\includegraphics[width=0.33\linewidth]{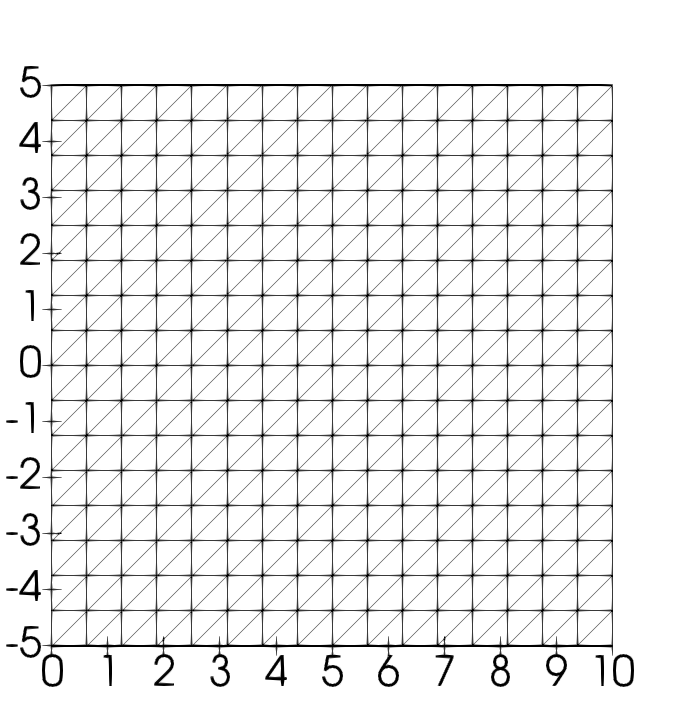}}
  \hfill
  \subfloat[]{\includegraphics[width=0.33\linewidth]{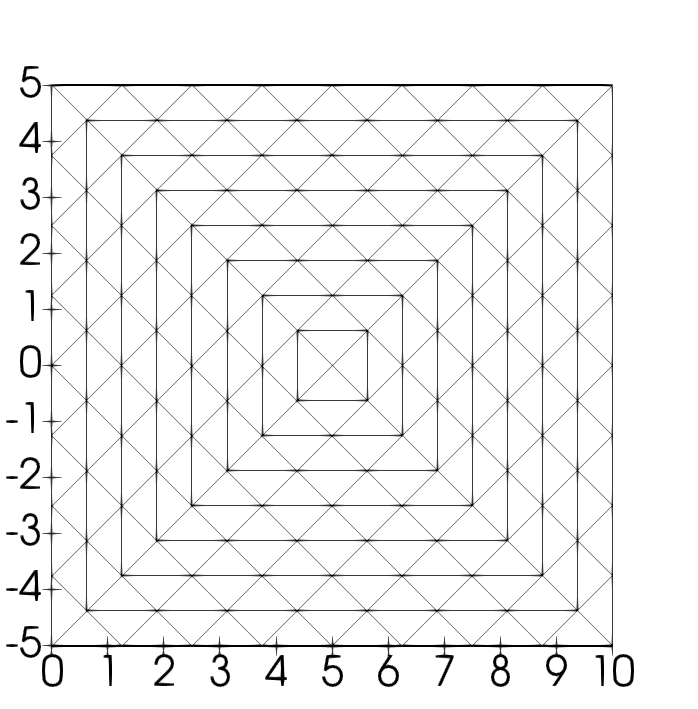}}
  \hfill
  \subfloat[]{\includegraphics[width=0.33\linewidth]{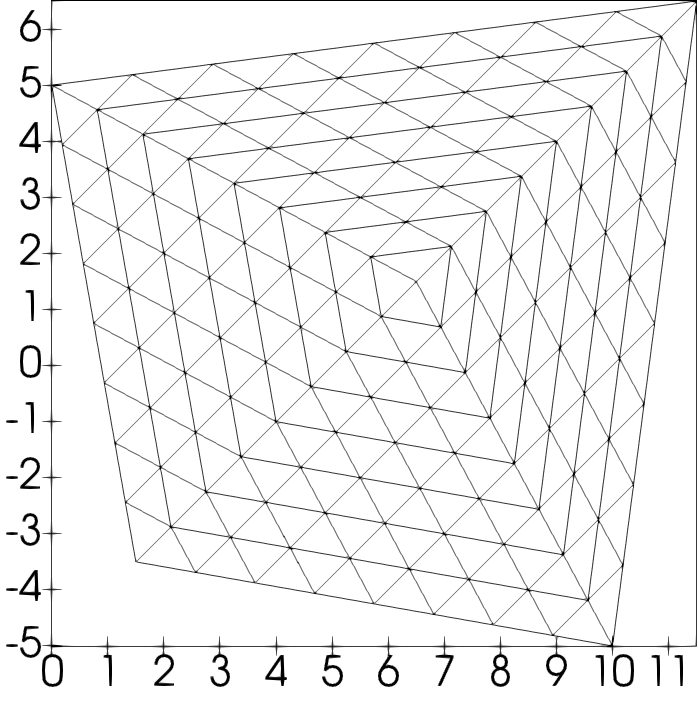}}
  \hfill
 \caption{Three types of grids used in the tests.}
  \label{fig:three_mesh}
\end{figure}

Numerical tests are performed on multiple linear and non-linear scalar
equations with diffusion and the Navier-Stokes equations using $P1$,
$P2$, and $P3$ nodal DG algorithms with the aRDG method.  Three types
of grids, as shown in Figure \ref{fig:three_mesh}, are
tested. Grid-\textit{a} and -\textit{b} are $0 \leq x \leq
10$. Grid-\textit{b} has the bottom-left corner moved to $(1.5,-3.5)$,
the top-right corner moved to $(11.5, 6.5)$, and the center moved to
$(6.5, 1.5)$. In grid-\textit{a}, each quadrilateral combined by two
adjacent triangles is a parallelogram, thus no error associated with
area truncation will be generated through the reconstruction
process. In grid-\textit{b}, large area truncation will occur on the
diagonals of the domain, where the combination of two adjacent
triangles forms a larger triangle with a larger area than the enclosed
parallelogram on which the reconstruction is performed.  In
grid-\textit{c}, the bottom-left and top-right corners are moved so
that larger area truncation to obtain an enclosed parallelogram for
reconstruction will occur along the top-left, top-right, and
bottom-right half of the diagonals.  However, the size of each element
is the same even though the shape is different. Among the four
sections of the diagonals in grid~\textit{c}, the top-right section
has the largest truncated area when obtaining an enclosed
parallelogram for reconstruction. Convergence studies are performed on
a series of systematic refinements of these three grids. Series of
grid-\textit{a} has 32, 128, 512, 2048, and 8192 elements, while
series of grid-\textit{b} and -\textit{c} have 16, 64, 256, 1024, 4096
elements.

In this section, the global $L_2$ and $L_\infty$ norms of the error are calculated as follows,
\begin{equation}
  L_2= \sqrt{ \frac{\sum_{k=1}^K \int_{\Omega_k}  \left[ u^k - u_e  \right]^2 d \Omega} { \sum_{k=1}^K|\Omega_k| }},
  \label{equ:L2_norm_cellUnit}
\end{equation}

\begin{equation}
  L_\mathrm{inf} = \max_{k=1}^K \frac{ \int_{\Omega_k}  \left|u^k - u_e \right|  d \Omega} { |\Omega_k|},
  \label{equ:Linf_norm_cellUnit}
\end{equation} 
where $u_e$ is the analytical solution. It is important to point out that
the errors calculated in this section contain both the spatial and temporal
discretization errors. Based on \citep{oberkampf2010verification}, the
error norms are,
\begin{equation}
  \left\|\varepsilon^{h_t}_{h_{{x}}}\right\| = g_x h_x^{\hat{p}} + g_t h_t^{\hat{q}}
  \label{equ:error_mixedOrder}
\end{equation}
where $g_x$ and $g_t$ are constants. $h_x$ is the spatial grid size
and $h_t$ is time-step size. For all the simulations presented in this
section, the five-stage fourth-order Runge-Kutta scheme
\citep{carpenter1994fourth} is used. The time step $h_t$ is calculated
from the most restrictive mesh refinement level and is fixed for all
meshes. When $h_t$ is fixed, equation \ref{equ:error_mixedOrder}
becomes,
\begin{equation}
  \left\|\varepsilon^{h_t}_{h_{{x}}}\right\| = g_x h_x^{\hat{p}} + \phi,
  \label{equ:error_mixedOrder_fixdt}
\end{equation}
where $\phi = g_t h_t^{\hat{q}}$ is the fixed temporal error
term. Then $\hat{p}$ can be evaluated with three mesh refinement
levels, e.g. coarse($r_x^2h_x$), medium($r_xh_x)$, and fine($h_x$),
\begin{equation}
  \hat{p} =  \frac{\ln\left(\frac{\left\|\varepsilon^{h_t}_{r^2_{{x}}h_{{x}}}\right\|-\left\|\varepsilon^{h_t}_{r_{{x}}h_{{x}}}\right\|}{\left\|\varepsilon^{h_t}_{r^2_{{x}}h_{{x}}}\right\|-\left\|\varepsilon^{h_t}_{h_{{x}}}\right\|}\right)}{\ln{(r_{{x}})}}.
  \label{equ:OOA_mixedOrder_fixdt}
\end{equation}

\subsection{Diffusion equation}
\begin{figure}[!htb]
 \centering
 \subfloat[t=0]{\includegraphics[width=0.45\linewidth]{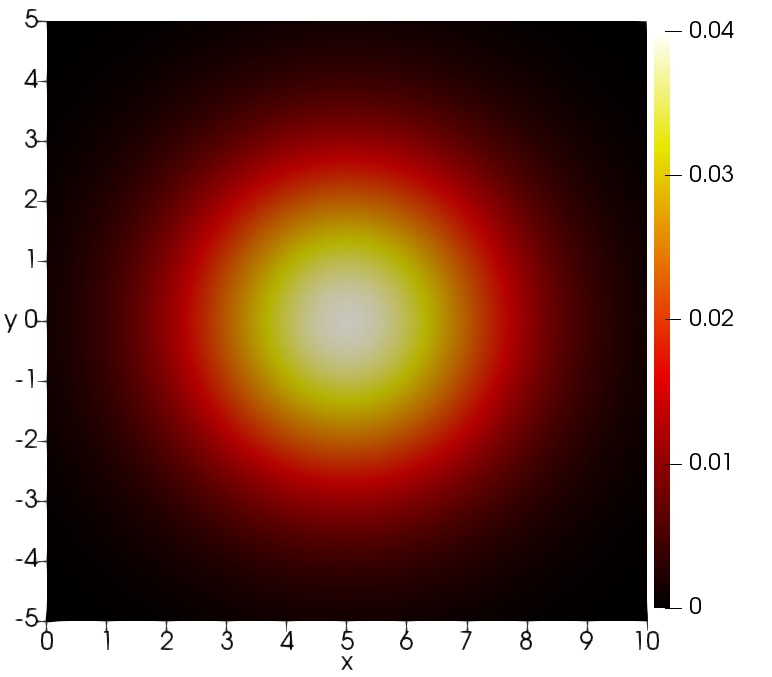}}
 \hfill
 \subfloat[t=0.5]{\includegraphics[width=0.45\linewidth]{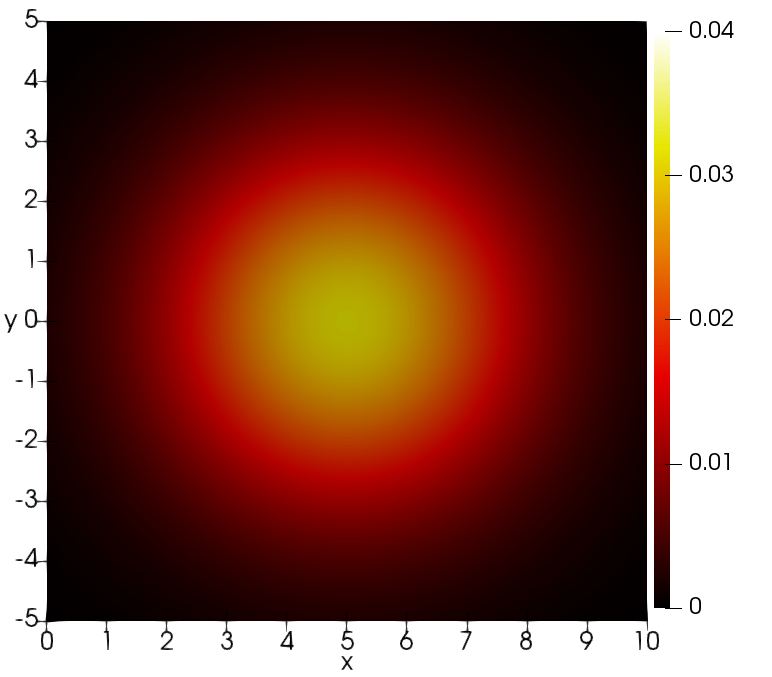}}
 \caption{Initial condition at $t=0$ and final solution at $t=0.5$ for the
   diffusion test. $P3$ test on grid-\textit{b} with 4096 elements is presented here.}
  \label{fig:diffusion_contour}
\end{figure}

The diffusion equation described in equation \ref{equ:diffusion} is
solved on the three grids presented in Figure \ref{fig:three_mesh}.
At $t=-D_0/D$, a solute of mass $M=1$ is loaded at $(x_0, y_0)$, where
$(x_0, y_0) =(5,0)$ for grid-\textit{a} and -\textit{b}, and $(x_0,
y_0) =(6.5,1.5)$ for grid-\textit{c}.  The analytical solution is
provided as
\begin{equation}
  u_e = \left( \frac{M}{4\pi (Dt+D_0)}\right) e^{-\frac{(x-x_0)^2 + (y-y_0)^2}{4(Dt+D_0)}} ,
  \label{equ:diffusion_exact}
\end{equation}
where $D=1$, and $D_0$ is set to be $2$ to make it numerically
feasible at $t=0$. This reconstruction follows equation
\ref{equ:NDG_diffusion_once_matrixForm_reference}, as the diffusion
coefficient is a constant. The initial condition at $t=0$ and final
solution of $t=0.5$ are presented in Figure
\ref{fig:diffusion_contour}.

\begin{figure}[!htb]
 \centering
 \includegraphics[width=\linewidth]{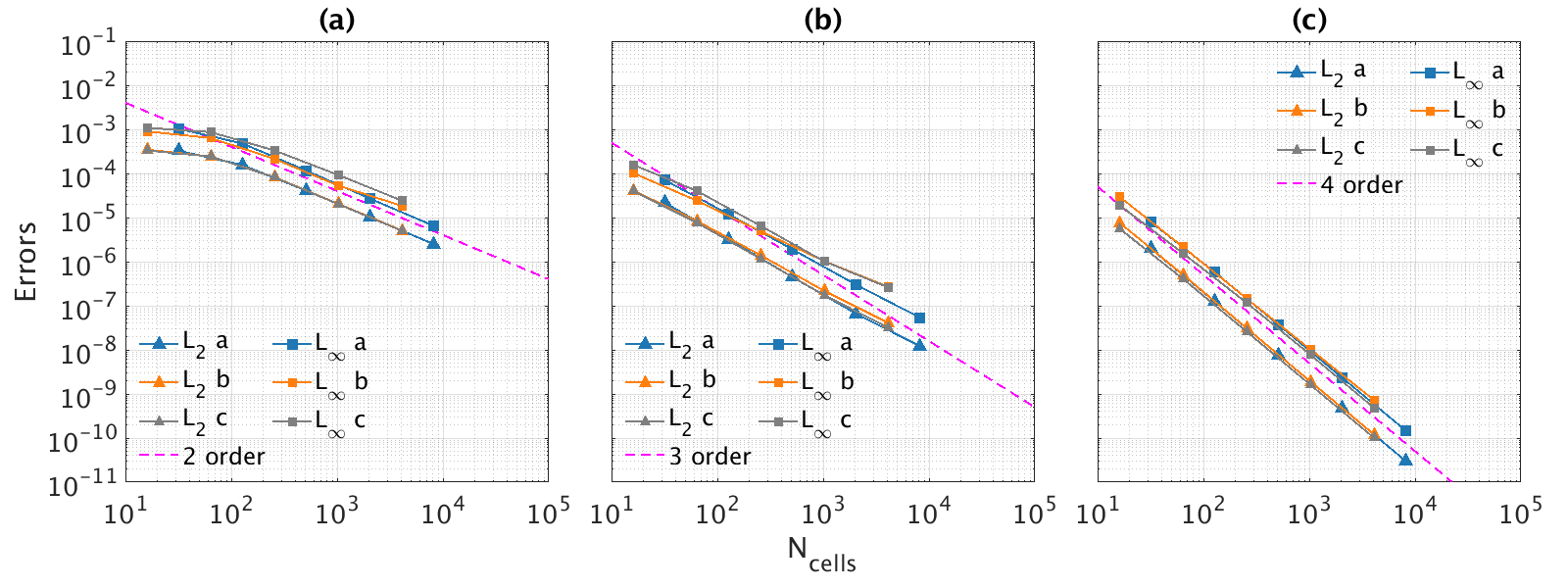}
 \caption{Convergence tests of the diffusion equation on three types
   of grids (Figure \ref{fig:three_mesh}) using (a) $P1$, (b) $P2$,
   and (c) $P3$ NDG algorithms. Formal orders of accuracy are
   indicated by the slopes with magenta lines.}
  \label{fig:diffusion_convergence}
\end{figure}

Results of the convergence study are presented in Figure
\ref{fig:diffusion_convergence}. The observed orders of accuracy are 
tabulated in Table \ref{table:purediff_ooa}. Both the convergence rates
of the $L_2$ and $L_\infty$ of errors for all three types of grids are
close to the formal order of accuracy $\hat{P} = P+1$
\citep{NDGbook_2007} for $P1$, $P2$, and $P3$ tests. The fact that
convergence lines of grid-\textit{a}, -\textit{b}, and -\textit{c} are
close to each other also indicates that the area truncation in the
aRDG process has minor impact on the accuracy of the scheme.  When two
triangles form a parallelogram, the density of degrees of freedom of
the reconstructed solution remains the same.  When the enclosed
parallelogram truncates a large area from the original adjacent
triangles that form a quadrilateral, the density of degrees of freedom
in the enclosed parallelogram is increased, which could compensate for
errors associated with the area truncation.

\begin{table}[!htbp]
  \centering
  \begin{tabular}{c|c c c | c c c}
    \toprule
    & &$L_2$ &  & &$L_\infty$ & \\   
    Mesh & $P1$ & $P2$ & $P3$ & $P1$ & $P2$ & $P3$ \\
    \midrule
    a & 2.015 & 2.828 & 4.005 & 2.123 & 2.707 & 3.988\\
    b & 2.003 & 2.754 & 4.018 & 2.216 & 2.347 & 3.819\\
    c & 2.014 & 2.802 & 4.005 & 1.812 & 2.850 & 3.908\\
    \bottomrule
  \end{tabular} 
  \caption{Observed orders of accuracy calculated from results presented in Figure \ref{fig:diffusion_convergence}}
  \label{table:purediff_ooa}
\end{table}

\subsection{Scalar advection-diffusion equation}
In order to test how well the aRDG diffusion solver couples with the
well-benchmarked NDG hyperbolic solver, this test focuses on the scalar
advection-diffusion equation,
\begin{equation}
  \frac{\partial u}{\partial t} + \vec{a} \cdot \nabla u - D \nabla^2
  u = 0 .
  \label{equ:advection_diffusion}
\end{equation}
The analytical solution is given by,
\begin{equation}
  u_e = \left( \frac{M}{4\pi (Dt+D_0)}\right) e^{-\frac{(x-a_xt-x_0)^2 +
      (y-a_yt-y_0)^2}{4(Dt+D_0)}}.
  \label{equ:advection_diffusion_exact}
\end{equation}
Similar to the diffusion test, equation
\ref{equ:NDG_diffusion_once_matrixForm_reference} is applied for the
reconstruction of the diffusion term here. A solute of mass is loaded
at $(x_0, y_0)$ at $ t=-D_0/D$, with $D=1$ and $D_0=2$. However,
$(x_0, y_0)$ is set to be $(4,-1.0)$ for all three types of grids
(Figure \ref{fig:three_mesh}), and a constant advection speed $\vec{a}
= (6,6)$ is chosen so that the diffusive mass is traveling along the
diagonal of the domain where truncation of area occurs in aRDG for
grid-\textit{b} and -\textit{c}. This way, the $L_\infty$ of the error
captures the error associated with area truncation in aRDG, if any.

\begin{figure}[!htb]
 \centering
 \subfloat[t=0]{\includegraphics[width=0.45\linewidth]{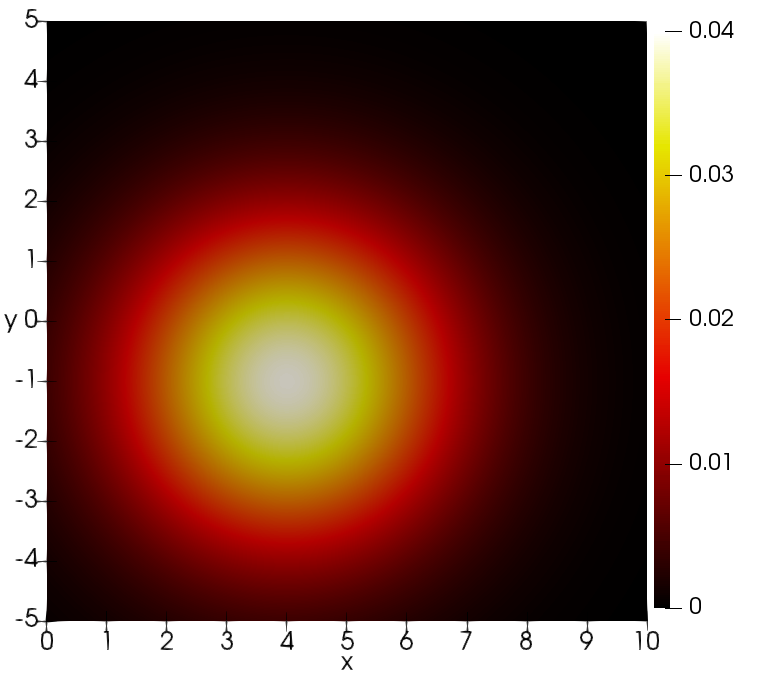}}
 \hfill
 \subfloat[t=0.5]{\includegraphics[width=0.45\linewidth]{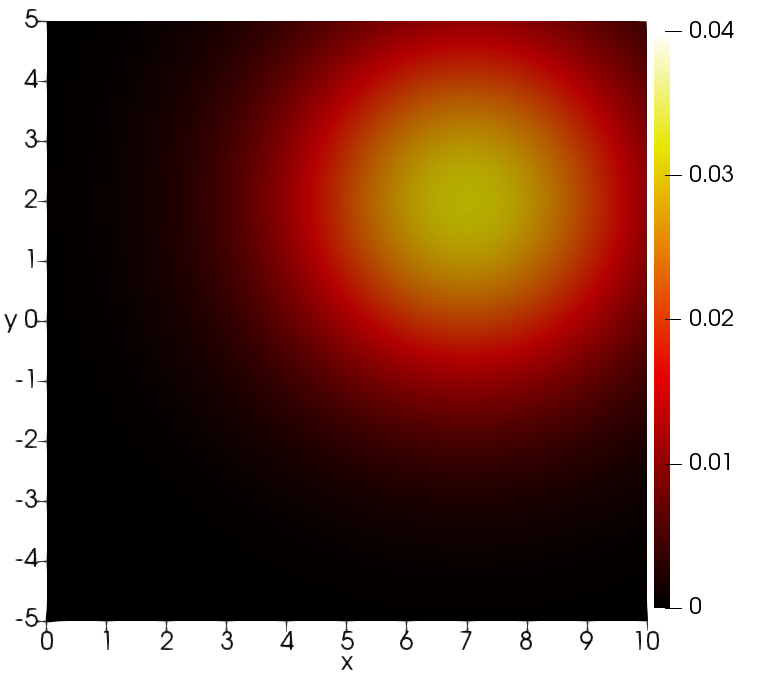}}
 \caption{Initial condition at $t=0$ and final solution at $t=0.5$ for the
   advection-diffusion test. $P3$ test on grid-\textit{b} with 4096
   elements is shown here.}
  \label{fig:adv_diff_contour}
\end{figure}

\begin{figure}[!htb]
 \centering
 \includegraphics[width=\linewidth]{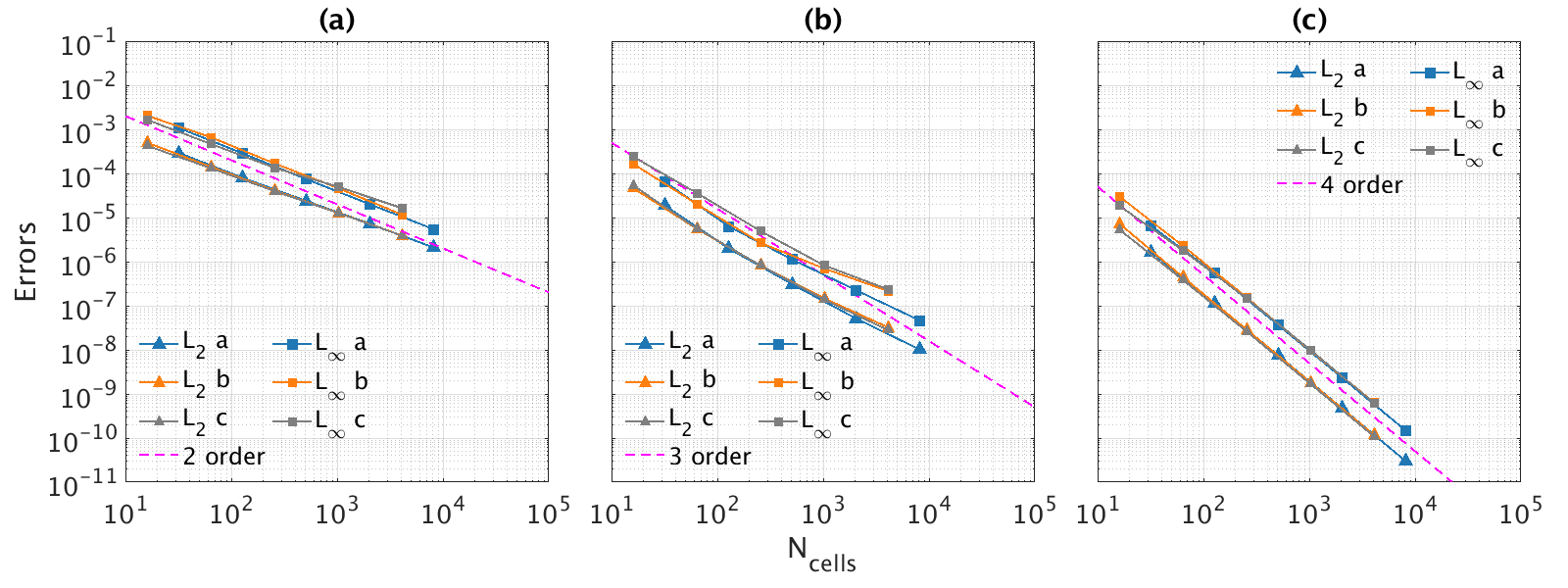}
 \caption{Convergence tests of advection-diffusion equation on three
   types of grids (Figure \ref{fig:three_mesh}) using (a) $P1$, (b) $P2$, and (c) $P3$ NDG
   algorithms. Formal orders of accuracy are indicated by the slopes with magenta lines.}
  \label{fig:adv_diff_convergence}
\end{figure}

The initial condition at $t=0$ and the final solution at $t=0.5$ are
presented in Figure \ref{fig:adv_diff_contour}. Convergence tests are
shown in Figure \ref{fig:adv_diff_convergence}. Similar to the pure
diffusion test case, the optimal convergence is achieved for all types
of meshes and polynomial orders that are tested. Again, the
convergence lines for all three grids are close to each other.

\begin{table}[!htbp]
  \centering
  \begin{tabular}{c|c c c | c c c}
    \toprule
    & &$L_2$ &  & &$L_\infty$ & \\   
    Mesh & $P1$ & $P2$ & $P3$ & $P1$ & $P2$ & $P3$ \\
    \midrule
    a & 1.690 & 2.629 & 3.970 & 1.897 & 2.321 & 3.961 \\
    b & 1.641 & 2.546 & 3.968 & 1.821 & 2.023 & 4.013 \\
    c & 1.603 & 2.638 & 3.936 & 1.344 & 2.815 & 3.873 \\
    \bottomrule
  \end{tabular} 
  \caption{Observed orders of accuracy calculated from results presented in Figure \ref{fig:adv_diff_convergence}}
  \label{table:adv_diff_ooa}
\end{table}

\subsection{Convection-diffusion equation with non-constant coefficients}
In order to test the robustness of the aRDG scheme on non-linear
equations, a scalar convection-diffusion equation with spatially- and
temporally-varying coefficients is employed here,
\begin{equation}
  \frac{\partial C}{\partial t} + \frac{1}{2}\frac{\partial}{\partial x} \left(a_{0x}CC\right) + \frac{1}{2}\frac{\partial}{\partial y} \left(a_{0y}CC\right)  -  \frac{\partial}{\partial x} \left(D_{0x}C \frac{\partial C}{\partial x} \right) - \frac{\partial}{\partial y} \left(D_{0y}C \frac{\partial C}{\partial y} \right) = S_{\textrm{MMS}} ,
  \label{equ:advection_diffusion_nonlinear}
\end{equation}
where $(a_{0x},a_{0y})$ and $(D_{0x},D_{0y})$ are constants. The
advection and diffusion coefficients are non-constant and do not
assume isotropicity. Equation
\ref{equ:NDG_diffusion_once_matrixForm_reference_general} is applied
here for the reconstruction of the diffusion terms.  The analytical
solution is constructed by method of manufactured solutions (MMS)
\citep{oberkampf2010verification}, a standard method used for code
verification.

\begin{figure}[!htb]
  \centering
  \includegraphics[width=\linewidth]{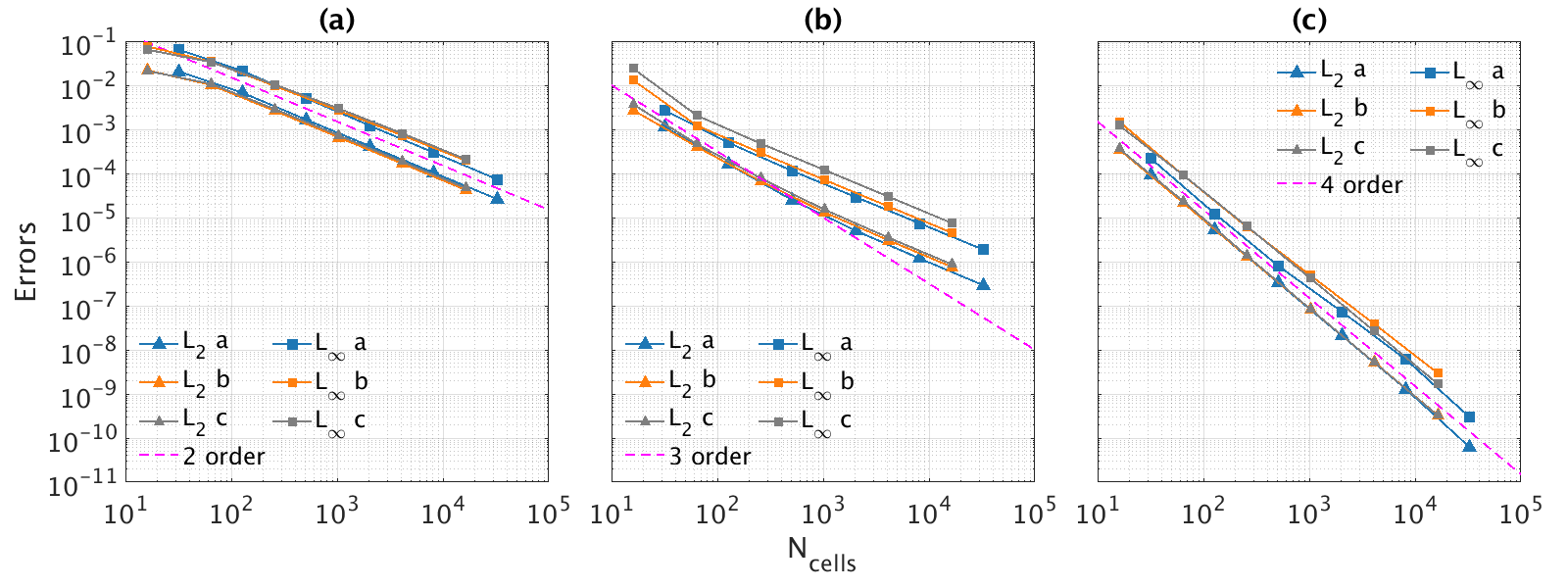}
  \caption{Convergence tests of scalar convection-diffusion equation
    with spatially- and temporally-varying coefficients on three types
    of grids (Figure \ref{fig:three_mesh}) using (a) $P1$, (b) $P2$,
    and (c) $P3$ NDG algorithms. Formal orders of accuracy are
    indicated by the slopes with magenta lines.}
  \label{fig:conv_diff_MMS_convergence}
\end{figure}

The results of the convergence tests are presented in Figure
\ref{fig:conv_diff_MMS_convergence}. The convergence rates agree with
the theoretical rates except for $P2$, where the observed rate is
slightly lower than the theoretical rates. This behavior is consistent
with previous results \citep{oden1998discontinuoushpfinite}.  Similar
to the linear test cases presented, no significant difference is found
between the results on different grids, which indicates that the
truncation of the area to obtain an enclosed parallelogram for
reconstruction does not introduce noticeable error into this system.

\begin{table}[!htbp]
  \centering
  \begin{tabular}{c|c c c | c c c}
    \toprule
    & &$L_2$ &  & &$L_\infty$ & \\   
    Mesh & $P1$ & $P2$ & $P3$ & $P1$ & $P2$ & $P3$ \\
    \midrule
    a & 1.993 & 2.113 & 3.974 & 2.034 & 1.971 & 3.525 \\
    b & 1.983 & 2.126 & 3.995 & 1.869 & 1.985 & 3.644 \\
    c & 1.982 & 2.136 & 4.011 & 1.867 & 2.004 & 3.970 \\
    \bottomrule
  \end{tabular} 
  \caption{Observed orders of accuracy calculated from results presented in Figure \ref{fig:conv_diff_MMS_convergence}}
  \label{table:conv_diff_MMS_ooa}
\end{table}

\subsection{Shear diffusion equation with non-constant coefficients}
\begin{figure}[!htb]
  \centering
  \includegraphics[width=\linewidth]{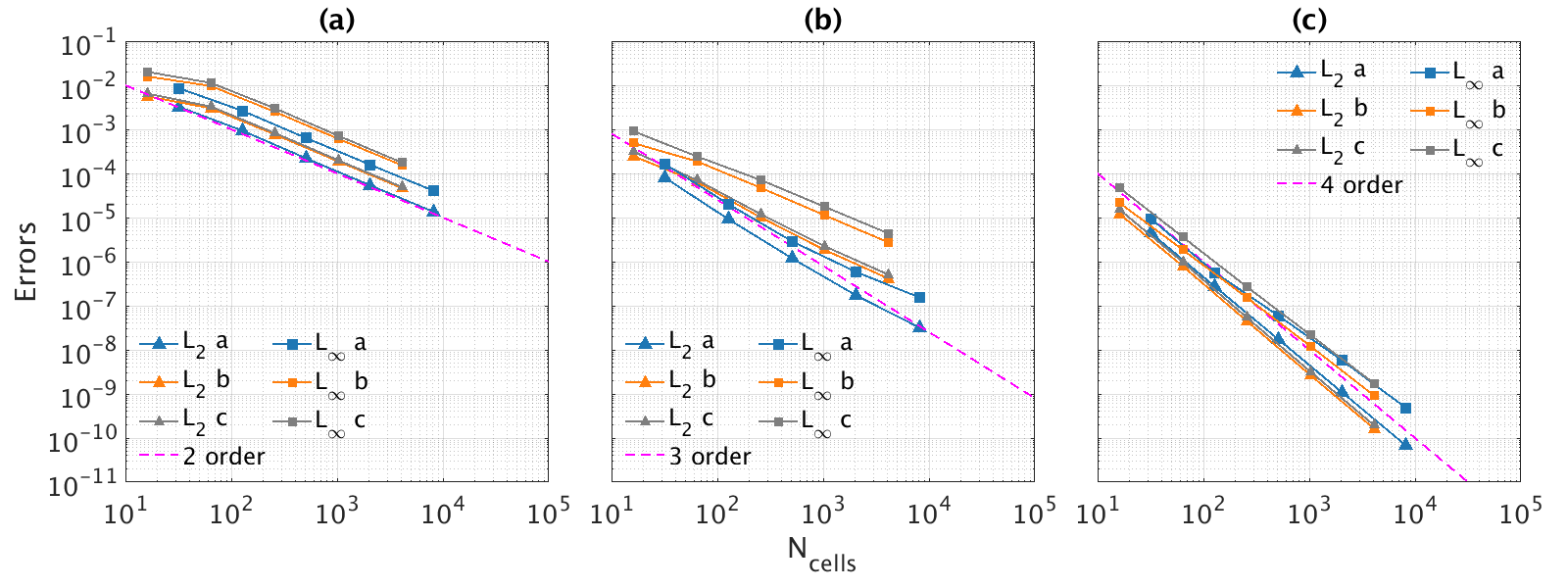}
  \caption{Convergence tests of scalar shear-diffusion equation with spatial 
   and temporal varying coefficients on three
   types of grids (Figure \ref{fig:three_mesh}) using (a) $P1$, (b) $P2$, and (c) $P3$ NDG
   algorithms. Formal orders of accuracy are indicated by the slopes with magenta lines.}
  \label{fig:shear_diff_MMS_convergence}
\end{figure}

Tests are performed on three types of grids (described in Figure
\ref{fig:three_mesh}) using the shear term in the diffusion equation
to further benchmark the robustness of aRDG algorithm.  Following the
work of \citep{johnson2019compact}, the shear diffusion equation is
described as,
\begin{equation}
  \frac{\partial C}{\partial t} 
  - \frac{\partial}{\partial x} \left(D_{0}C \frac{\partial C}{\partial x} \right) - \frac{\partial}{\partial y} \left(D_{0}C \frac{\partial C}{\partial y} \right)
  -  \theta D_{0} \left[ \frac{\partial}{\partial x} \left(C \frac{\partial C}{\partial y} \right) + \frac{\partial}{\partial y} \left(C \frac{\partial C}{\partial x} \right)  \right]  = S_{\textrm{MMS}} ,
  \label{equ:diffusion_shear}
\end{equation}
where $\theta=\frac{1}{6}$. Equation
\ref{equ:NDG_diffusion_once_matrixForm_reference_general} is applied
here for the reconstruction of the diffusion term, and the convergence
results are presented in Figure
\ref{fig:shear_diff_MMS_convergence}. In this study, noticeable
differences in the convergence errors from three types of grids can be
observed on $P1$ and $P2$ tests. Convergences rates agree well with
theory, except in $P2$ tests, where the convergence rates on
grid-\textit{b} and grid-\textit{c} are slower than the theoretical
rate.  The accuracy of aRDG appears to be more sensitive to area
truncation necessary to obtain the enclosed parallelogram for $P2$
shear diffusion problems. However, no significant difference can be
observed on different grids for $P3$ tests, and the computed
convergence rates successfully predict the theory.  

\begin{table}[!htbp]
  \centering
  \begin{tabular}{c|c c c | c c c}
    \toprule
    & &$L_2$ &  & &$L_\infty$ & \\   
    Mesh & $P1$ & $P2$ & $P3$ & $P1$ & $P2$ & $P3$ \\
    \midrule
    a & 2.029 & 2.843 & 3.992 & 2.008 & 2.403 & 3.354 \\
    b & 2.053 & 2.513 & 4.075 & 2.064 & 2.075 & 3.653 \\
    c & 2.069 & 2.487 & 4.089 & 2.099 & 2.005 & 3.637 \\
    \bottomrule
  \end{tabular} 
  \caption{Observed orders of accuracy calculated from results presented in Figure \ref{fig:shear_diff_MMS_convergence}}
  \label{table:shear_diff_MMS_ooa}
\end{table}

The algorithms presented here and in \citep{johnson2019compact}
exclusively use face neighbors of the element to perform the
reconstruction and recovery, respectively.  For complete consistency
with accurately resolving the shear term in the diffusion equation,
particularly as the shear term becomes significant, it may be necessary to
account for all face and vertex neighbors of the elements.  However, a
practical implementation including all vertex neighbors while
maintaining computational and storage efficiency is non-trivial for
unstructured grids and is a subject of future work.  The likely reason
that the shear term here still produces sufficient order of accuracy
is due to (i) the normal stresses being dominant as is the case in
most physical systems and (ii) the fourth-order Runge-Kutta
time-integration scheme sufficiently resolving the cross derivatives
over the five stages for the problems tested.

\subsection{Navier-Stokes equations}
This section applies the aRDG algorithm to the compressible
Navier-Stokes equations,
\begin{equation}
  \frac{\partial \bm{Q}}{\partial t} + \frac{\partial \bm{F}_i}{\partial x_i} + \frac{\partial \bm{G}_i}{\partial x_i} =0 ,\quad i = 1, \dots, N_d,
  \label{eq:aRDG_adv_diff_general}
\end{equation}
where
\begin{gather}
  \bm{Q} =
  \begin{pmatrix}
    \rho \\
    \rho u_j\\
    \epsilon
  \end{pmatrix}
  ,\quad
  \bm{F}_i =
  \begin{pmatrix}
    \rho u_i \\
    \rho u_iu_j+p\delta_{ij}\\
    (\epsilon + p) u_i
  \end{pmatrix}
  ,\quad
  \bm{G}_i =
  \begin{pmatrix}
    0\\
    -\Pi_{ij}\\
    -u_j\Pi_{ij}+q_i
  \end{pmatrix} ,
  \label{equ:aRDG_Navier_Stokes}
\end{gather}
and the viscous stress tensor $\Pi$ and heat flux $q$ are given by 
\begin{gather}
    \Pi_{ij} = \mu \left( \frac{\partial u_i}{\partial x_j} + \frac{\partial u_j}{\partial x_i} \right)
    - \frac{2}{3}\mu\nabla\cdot\bm{u}\delta_{ij} ,     \label{equ:NS_viscosity} \\
    q_i = -\kappa \frac{\partial T}{\partial x_i}.
    \label{equ:NS_heatflux}
\end{gather}
The molecular viscosity $\mu$ is calculated through Sutherland's law \citep{sutherland1893lii} and thermal conductivity $\kappa$ is calculated as 
\begin{equation}
  \kappa = \frac{C_p \mu}{Pr},
\end{equation}
where the Prandtl number $Pr$ is $0.7$.

Two sets of tests are performed. The first one is a code verification test and
the second one is a model validation test.
\subsubsection{Method of Manufactured Solutions (MMS)}
\begin{figure}[!htb]
 \centering
 \includegraphics[width=\linewidth]{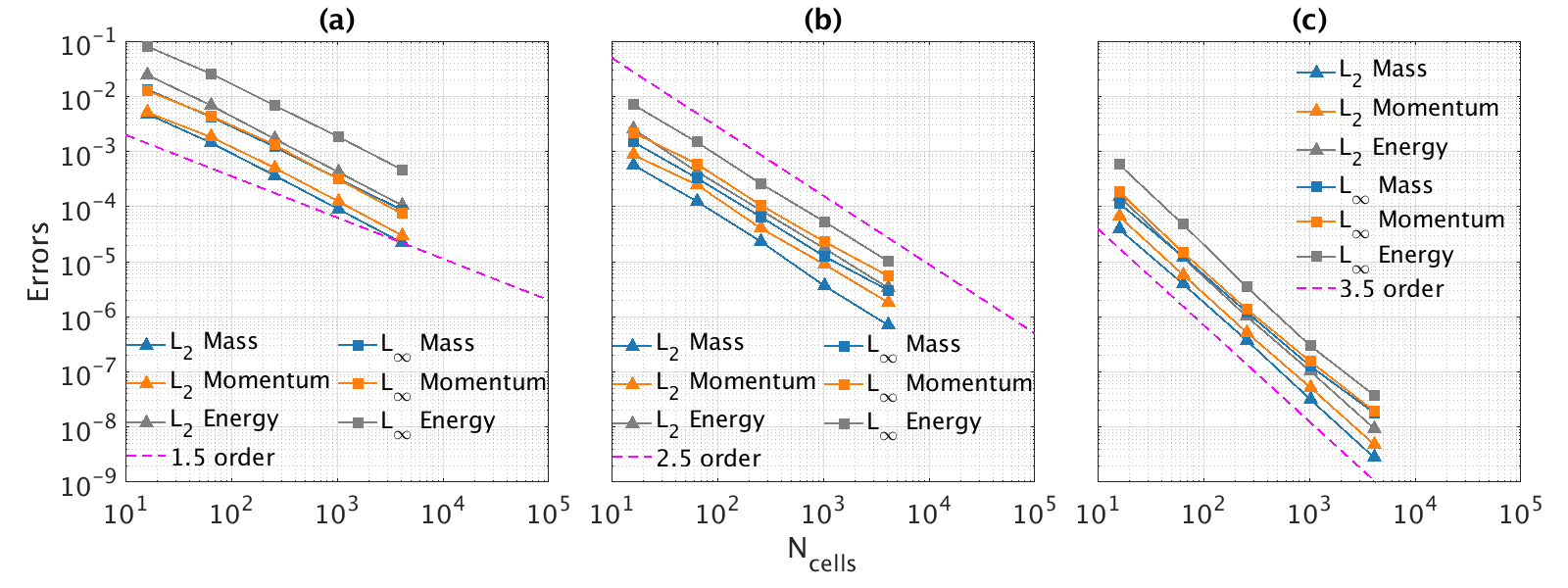}
 \caption{Convergence tests of compressible Navier-Stokes equations on 
   grid-\textit{b} (Figure \ref{fig:three_mesh}) using (a) $P1$, (b) $P2$, and (c) $P3$ NDG
   algorithms. Convergence rates for mass, momentum, and total energy are presented.
   Formal orders of accuracy are indicated by the slopes with magenta lines.}
  \label{fig:NavierStokes_MMS}
\end{figure}

Code verification is performed on grid-\textit{b} (Figure
\ref{fig:three_mesh}) using MMS. Lax-Friedrichs
\citep{toro2013riemann} flux is applied here for the hyperbolic
terms. According to \citep{NDGbook_2007}, the optimal order of
accuracy of the NDG algorithm for a system is $P+1/2$, when a general
monotone flux is used. The results are presented in Figure
\ref{fig:NavierStokes_MMS}. The observed orders of accuracy for all
three variables in $P1$ tests are slightly higher than the optimal
rate. Results of $P2$ and $P3$ tests show good agreement with theory.

\begin{table}[!htbp]
  \centering
  \begin{tabular}{c|c c c | c c c}
    \toprule
    & &$L_2$ &  & &$L_\infty$ & \\   
    Variable & $P1$ & $P2$ & $P3$ & $P1$ & $P2$ & $P3$ \\
    \midrule
    Mass & 2.027 & 2.736 & 3.575 & 1.930 & 2.484 & 3.323 \\
    Momentum & 2.024 & 2.143 & 3.310 & 2.080 & 2.238 & 3.168 \\
    Energy & 2.004 & 2.289 & 3.294 & 1.853 & 2.270 & 3.624 \\
    \bottomrule
  \end{tabular} 
  \caption{Observed orders of accuracy calculated from results presented in Figure \ref{fig:NavierStokes_MMS}}
  \label{table:NavierStokes_MMS_ooa}
\end{table}

\subsubsection{Flow over cylinder}
\begin{figure}[!htb]
 \centering
 \includegraphics[width=\linewidth]{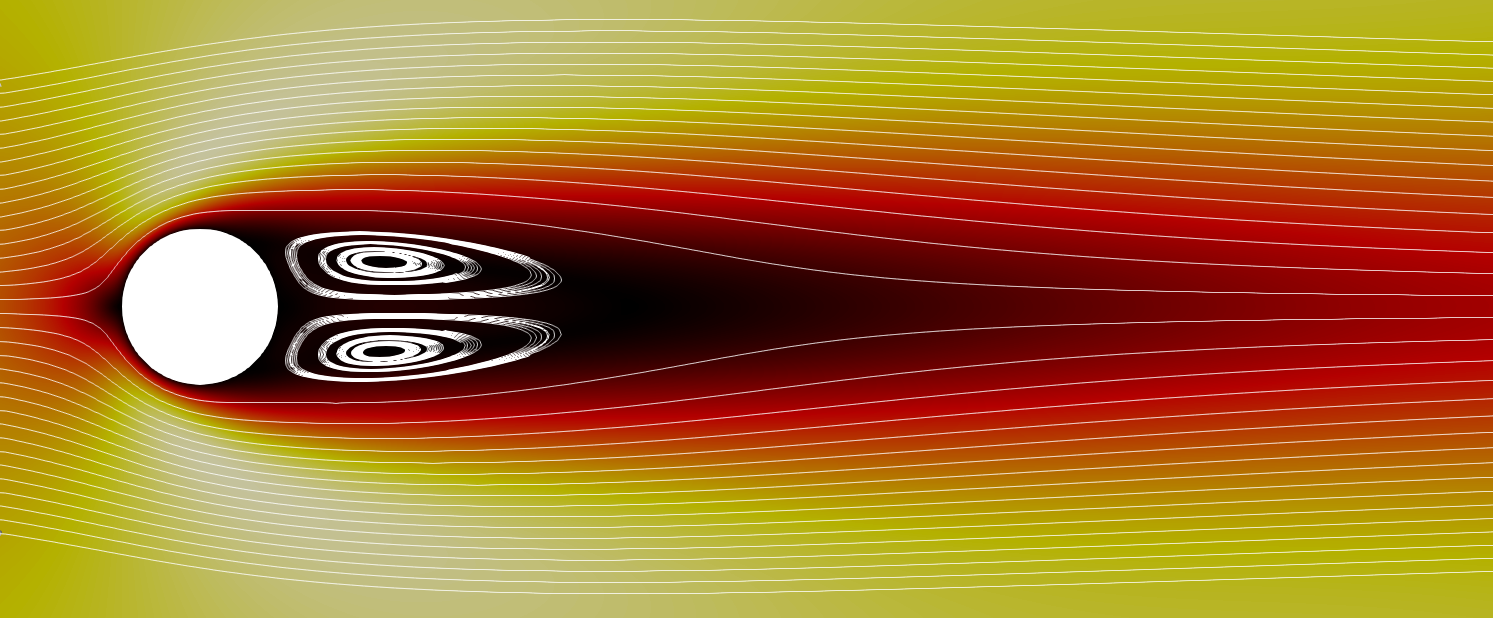}
   \caption{Mach number plot with streamlines of subsonic flow over
     circular cylinder with $Re=40$}
  \label{fig:FOC_Re40_Mach}
\end{figure}

Model validation is performed on an subsonic flow over cylinder case
with $Re = 40$. A circular cylinder with a diameter of $D$ is placed
at the center of a domain of size $32D \times 16D$. The computed Mach
number is plotted in Figure \ref{fig:FOC_Re40_Mach} with streamlines
indicating the recirculation. The drag coefficient and the length of
the recirculation region are calculated and presented in Table
\ref{table:FOC_Re40}, which agree well with \cite{tseng2003ghost}.

\begin{table}[h!]
\caption{Drag coefficient and length of recirculation for subsonic flow over circular cylinder with $Re=40$}
\begin{center}
\begin{tabular}{c|c|c}
\toprule
$Re=40$ & Drag coefficient & Length of recirculation \\
\midrule
Current study   & 1.47  & 2.26$D$ \\
Tseng and Ferziger \citep{tseng2003ghost}   & 1.53  & 2.21$D$ \\
\bottomrule
\end{tabular}
\end{center}
\label{table:FOC_Re40}
\end{table}

\subsubsection{High-energy-density implosion hydrodynamics}  \label{sec:ICF}
Numerical simulations of high-energy-density implosion hydrodynamics
relevant to inertial confinement fusion and astrophysics are
challenging due to the large gradients in density, temperature, and
pressure in these regimes that increase substantially as the
implosions progress in time.  While a number of 1-dimensional tools
exist that are able to access these regimes, multi-dimensional
simulations remain a challenge due to the growth of hydrodynamic
instabilities at the sharp interfaces \cite{clark2016three,
  srinivasan2012magnetic, srinivasan2012mechanism,
  wang2017theoretical, srinivasan2013mitigating, li2018richtmyer,
  srinivasan2014role}, the need to resolve general geometries by
mitigating the effects of grid shapes from affecting the dynamics
\cite{joggerst2014cross}, the highly disparate parameters that are
encountered across relatively short spatial scales
\cite{srinivasan2014mitigating}, and the need to resolve disparate
spatial and temporal scales, to name a few.  Furthermore, a majority
of numerical simulations do not incorporate the highly disparate
Reynolds numbers (and magnetic Reynolds numbers if including magnetic
fields using magnetohydrodynamic models) that occur in these regimes
\cite{srinivasan2014mitigating}.  To address these challenges, this
work demonstrates the application of the unstructured mesh aRDG
algorithm developed here for implosion simulations in
high-energy-density hydrodynamics employing highly disparate
densities, temperatures, and viscosities over short spatial scales.

The radial implosion problem setup \cite{song2020dissertation} is
adapted from \cite{joggerst2014cross}. In \cite{joggerst2014cross},
the circular shape of the implosion without any perturbation is well
maintained when using a spherical coordinate system. However, the
circular shape of the implosion is changed by a structured mesh in
Cartesian coordinate system, limiting the geometric flexibility of
both types of coordinate systems in these codes.  To explore this in
the unstructured DG code using the aRDG algorithm for diffusion,
simulations are performed on one quadrant of a circle.

\begin{figure}
  \centering
  \includegraphics[width=0.6\linewidth]{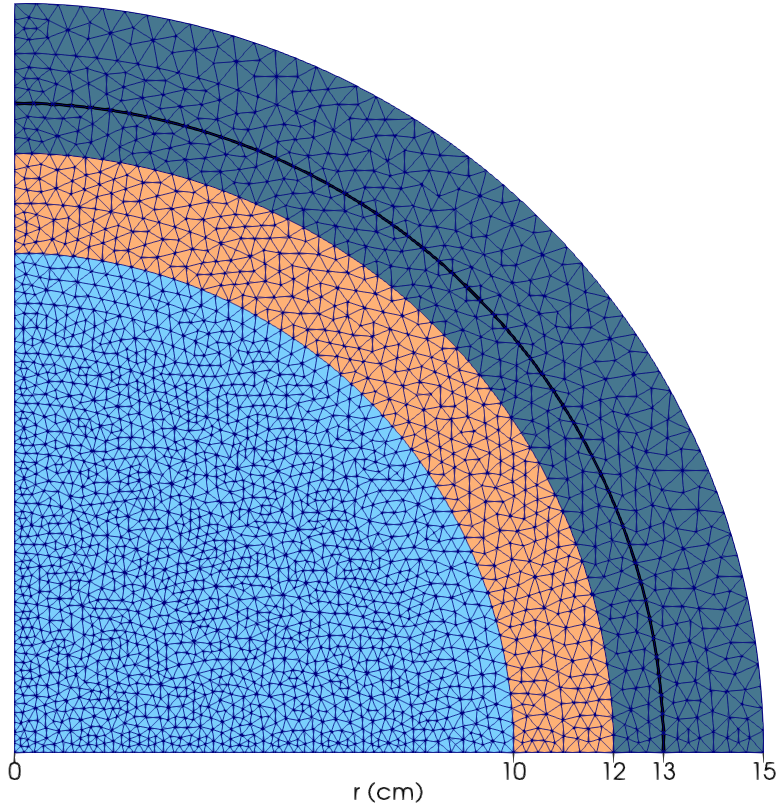}
  \caption{Illustration of the mesh used in radial implosion test with
    a coarse version}
  \label{fig:imp_mesh}
\end{figure}

Simulations are performed with an unstructured mesh of approximately
1,000,000 triangular elements.  The mesh elements are guided by a
series of circles with size of the element proportionally decreasing
moving inward in radius until a radius well within the inner fluid,
within which the element size remains similar.  A lower resolution
illustration of this mesh is presented in Figure \ref{fig:imp_mesh}.
The inner region for $r< \SI{10}{cm}$ in Figure \ref{fig:imp_mesh} is
a low-density region, followed by a high-density region for
$\SI{10}{cm} < r < \SI{12}{cm}$ with an Atwood number of $0.9$ across
the $r= \SI{10}{cm}$ interface. For $r > \SI{12}{cm}$ there is a
low-density, high-pressure region that acts as a pusher for the
implosion.  An initial random multimode perturbation is applied at the
interface between the inner region and the dense shell ($r=
\SI{10}{cm}$).

\begin{figure}
  \centering
  \includegraphics[width=0.9\linewidth]{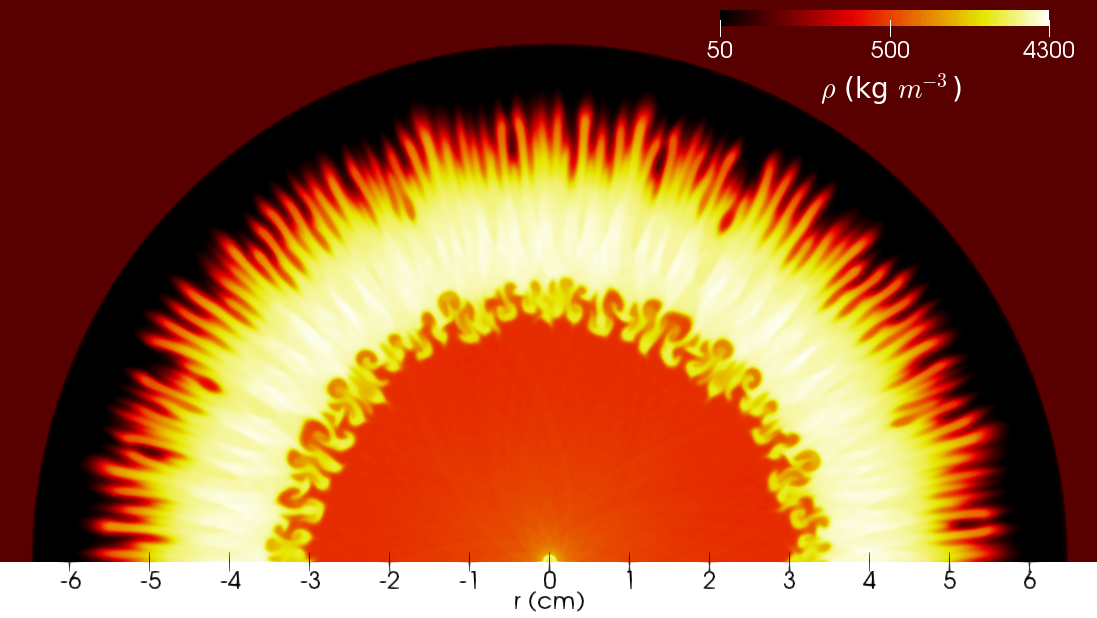}
  \caption{Density profile of an inviscid implosion case with
    multimode perturbation at time $t=\SI{2.5}{s}$ using the Euler
    equations. Mesh resolution: 1,000,000 triangular elements.  Note
    the growth of significant short-wavelength RTI.}
  \label{fig:imp1M_multi_novis_density_t25}
\end{figure}

The density profile at $t=\SI{2.5}{s}$ is presented in Figure
\ref{fig:imp1M_multi_novis_density_t25} for an inviscid
case. Significant turbulent mixing due to the growth of the
Rayleigh-Taylor instability (RTI) can be observed at the inner
interface. Note that the small scale features of the RTI mixing are
well captured even with these high density gradients.

\begin{figure}
  \centering
  \includegraphics[width=0.9\linewidth]{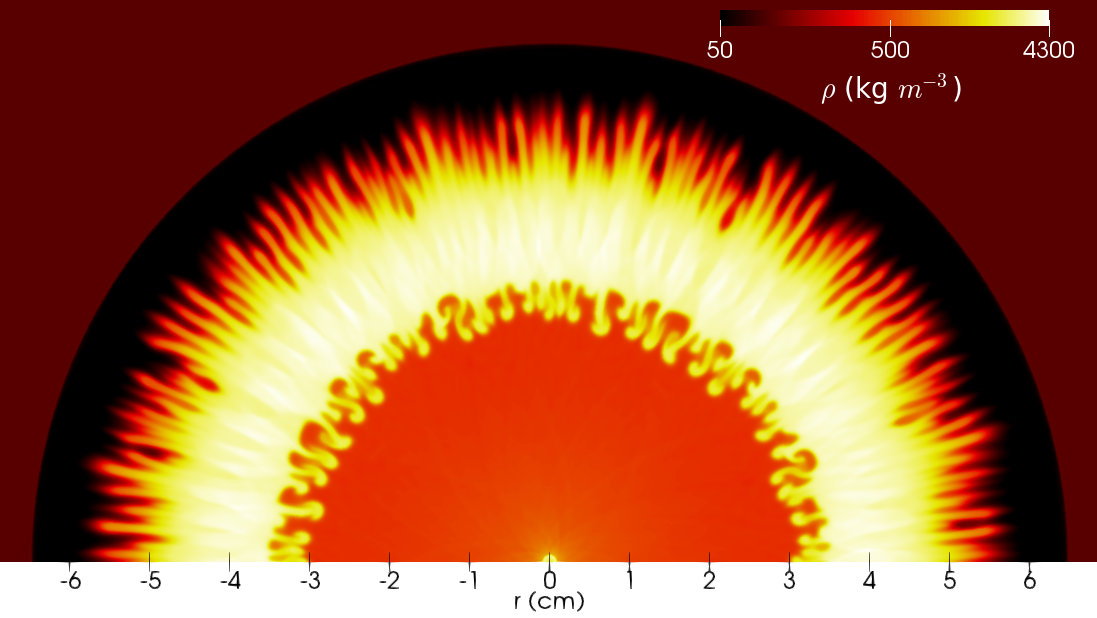}
  \caption{Density profile of implosion case with multimode
    perturbation at time $t=\SI{2.5}{s}$ using the Navier-Stokes
    equations, with Re$\sim 1,300$. Mesh resolution: 1,000,000
    triangular elements.  Note some stabilization of short-wavelength
    modes and decreased amplitude of the RTI. }
  \label{fig:imp1M_multi_mu1em4_density_t25}
\end{figure}

\begin{figure}
  \centering
  \includegraphics[width=0.9\linewidth]{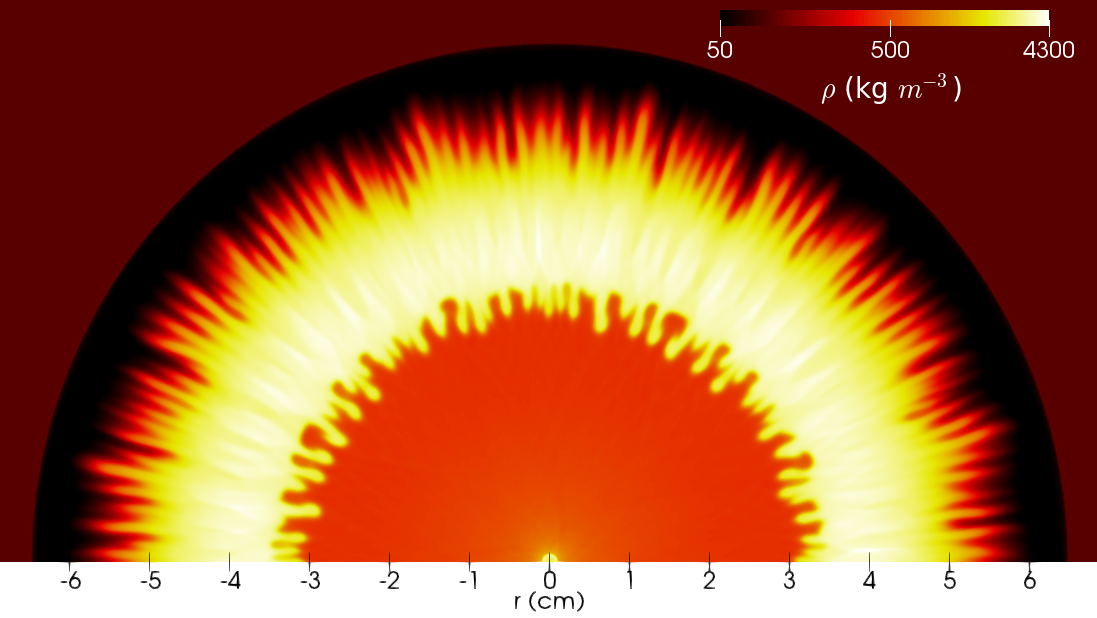}
  \caption{Density profile of implosion case with multimode
    perturbation at time $t=\SI{2.5}{s}$ using the Navier-Stokes
    equations, with Re$\sim 400$. Mesh resolution: 1,000,000
    triangular elements. Note significant stabilization of
    short-wavelength modes and decreased amplitude of the RTI. }
  \label{fig:imp1M_multi_mu3em4_density_t25}
\end{figure}

Simulations are performed applying the aRDG algorithm to include
disparate viscosities and explore their impact on the RTI growth
during implosions. An interface tracking algorithm is used to apply
the corresponding viscosity to the different sides of the interface,
thus accounting for disparate Reynolds numbers across a sharp
interface region.  The dense shell fluid viscosity corresponds to an
inviscid regime.  For the inner fluid, the viscosity is varied such
that Reynolds numbers of approximately $1,300$ and $400$ are
explored. The density evolution for these two cases at $t=\SI{2.5}{s}$
are presented in Figures \ref{fig:imp1M_multi_mu1em4_density_t25} and
\ref{fig:imp1M_multi_mu3em4_density_t25}, respectively.  RTI growth is
impacted by viscosity where an inviscid simulation would permit
development of turbulence while large viscosities stabilize short
wavelength modes adjusting the flow to be more laminar.  The effect of
disparate viscosity across an interface with RTI growth, where the
bubbles grow into inviscid regions while spikes grow into viscous
regions, constitutes open and important research in the field of
high-energy-density hydrodynamics.  While these simulations
sufficiently demonstrate the capability of the aRDG algorithm to
resolve disparate viscosities, even more extreme Reynolds number
variation across an interface will constitute future physics studies.

\section{Conclusion}
In this paper, an affine reconstructed discontinuous Galerkin method
has been described to solve the diffusion operator accurately and
efficiently on unstructured grids of triangles.  The algorithm is
demonstrated on a substantive problem in high-energy-density
hydrodynamics where disparate densities, pressures, and viscosities
present a significant challenge in effectively resolving radially
imploding dynamics and corresponding hydrodynamic instability
development.  A practical guideline on how to apply this algorithm to
the nodal discontinuous Galerkin method has been provided. All
computations can be done on the reference domain, which couples well
with the notable nodal discontinuous Galerkin scheme from
\citep{NDGbook_2007}. Benchmark tests are performed on three types of
grids with different refinement levels using $P1$, $P2$, and $P3$ NDG
schemes for linear and non-linear scalar equations with diffusion and
the Navier-Stokes equations.  The observed orders of accuracy
generally agree with the formal orders of accuracy for all tests.
Some $P2$ results have a $\mathcal{O}(h_x^2)$ convergence as described
in \citep{oden1998discontinuoushpfinite} which shows that the optimal
order of accuracy of DG for diffusion is $\mathcal{O}(h_x^{P+1})$ for
odd $P$ and $\mathcal{O}(h_x^{P})$ for even $P$.  By maintaining the
same polynomial order for the described reconstruction method, the
density of nodes in the reconstructed element on the physical domain
is not decreased, which means discretization error is not increasing
through this reconstruction. When two triangles form a parallelogram,
the density of degrees of freedom of the reconstructed solution
remains the same. When the enclosed parallelogram truncates a large
area from the original adjacent triangles that form a quadrilateral,
the density of degrees of freedom in the enclosed parallelogram is
increased, which could compensate for errors associated with the area
truncation. This may explain why the errors associated with all three
types of grids are very close to each other for most of the tests
presented, except for when the shear term is included in the
diffusion. It is also straightforward to extend the aRDG algorithm to
other types of elements as long as an enclosed parallelogram can be
found in adjacent elements. Future work will focus on extending the
aRDG algorithm to three dimensional unstructured grids.

\section*{Funding Sources}
This work was supported by the US Department of Energy under grant
number DE-SC0016515.

The author acknowledges Advanced Research Computing at Virginia Tech
for providing computational resources and technical support that have
contributed to the results reported within this work.  URL:
http://www.arc.vt.edu

\section*{References}
\bibliography{reference}

\end{document}